\documentclass[preprint2]{aastex}
\pdfoutput=1

\newcommand{\dthree}{D\ensuremath{_{3}}}
\newcommand{\heiv}{\ensuremath{\lambda}5876}
\newcommand{\heir}{\ensuremath{\lambda}10830}
\newcommand{\vsini}{\ensuremath{v\sin i}}
\newcommand{\BV}{\ensuremath{B-V}}
\newcommand{\Wobs}{\ensuremath{W_\mathrm{obs}}}
\newcommand{\Wmax}{\ensuremath{W_\mathrm{max}}}
\newcommand{\Wq}{\ensuremath{W_\mathrm{q}}}
\newcommand{\Wa}{\ensuremath{W_\mathrm{a}}}
\newcommand{\Teff}{\ensuremath{T_\mathrm{eff}}}

\slugcomment{Draft version \today}

\shorttitle{Area of Active Regions in Dwarf Stars}
\shortauthors{Andretta et al.}

\begin{document}

\title{Estimates of Active Region Area Coverage through Simultaneous
  Measurements of \ion{He}{1} $\lambda\lambda$ 5876 and 10830 Lines}

\author{Vincenzo Andretta\altaffilmark{1}}
\affil{INAF -- Osservatorio Astronomico di Capodimonte\\
  Salita Moiariello, 16\\
  80131 Naples, Italy\\
  andretta@oacn.inaf.it}
\author{Mark S. Giampapa}
\affil{National Solar Observatory\altaffilmark{2}\\
  950 N.~Cherry Avenue\\
  Tucson, AZ~85719, USA}
\author{Elvira Covino}
\affil{INAF -- Osservatorio Astronomico di Capodimonte\\
  Salita Moiariello, 16\\
  80131 Naples, Italy}
\author{Ansgar Reiners}
\affil{Institut f\"ur Astrophysik\\
  Georg-August-Universit\"at G\"ottingen\\
  Friedrich-Hund-Platz 1\\
  37077 G\"ottingen, Germany}
\and
\author{Benjamin Beeck\altaffilmark{1}}
\affil{Max Planck Institute for Solar System Research\\
  Justus-von-Liebig-Weg 3\\
  37077 G\"ottingen, Germany}

\altaffiltext{1}{Visiting Astronomer, European Southern Observatory}
\altaffiltext{2}{Operated by the Association of Universities for Research in Astronomy under a cooperative agreement with the National Science Foundation}

\begin{abstract}
  Simultaneous, high-quality measurements of the neutral helium triplet features
  at 5876~\AA\
  and 10830~\AA, respectively, in a sample of solar-type stars are
  presented. 
    The observations were made with ESO telescopes at the La Silla Paranal
    Observatory under program ID 088.D-0028(A) and MPG Utility Run for FEROS
    088.A-9029(A).
  The equivalent widths of these features combined with
  chromospheric models are utilized to infer the fractional area coverage, or
  filling factor, of magnetic regions outside of spots. We find that the
  majority of the sample is characterized by filling factors less than unity.
  However, discrepancies occur among the coolest K-type and warmest and
  most rapidly rotating F-type dwarf stars.  
    We discuss these apparently anomalous results and find that in the case
    of K-type stars they are an
    artifact of the application of chromospheric models best suited to the Sun
    than to stars with significantly lower \Teff.  The case of the F-type rapid
    rotators can be explained with the measurement uncertainties of the
    equivalent widths, but they may also be due to a non-magnetic heating 
    component in their atmospheres.
  With the exceptions noted above, preliminary
  results suggest that the 
    average heating rates in the active regions are the same from one star to
    the other, differing in the spatially integrated, observed level of
    activity due to the area coverage.
  Hence, differences in activity in this sample
  are mainly due to the filling factor of active regions. 
\end{abstract}


\section{Introduction}

While sunspots are the most visible manifestations of magnetic flux emergence
resulting from dynamo processes, magnetic flux concentrations outside of spots
          in active regions  form a significant fraction of the total
(unsigned) magnetic flux budget of the Sun.           Likewise, the total and
  spectral solar irradiance as functions of time cannot be modeled by
  considering the contribution of sunspots only \citep[for a recent review,
  see][]{yeo-etal:14}. 

Determining the distribution, or at least the fractional area coverage of
magnetic active regions, is relevant to both dynamo theory and to models of
chromospheric and coronal heating.  With regard to the latter, flux-calibrated
chromospheric emission line profiles yield the surface-averaged emission that
represents a lower limit to the intrinsic emission in localized active
regions.  A more accurate knowledge of the actual radiative losses resulting
from chromospheric heating would provide a further constraint for the
development of models based on, for example, local plasma heating by Joule
dissipation associated with an Alfv\'en wave field \citep{tu13}. While information on
the spatial distribution of magnetically active regions on stellar surfaces can be obtained in
some special cases (mostly rapid rotators through Doppler imaging), such
measurements have always been elusive in more solar-like stars.

A census of the solar magnetic flux in its
various forms can be performed directly with the distinct advantage of
spatially resolved observations.  In the case of stars, however, we generally
rely on radiative proxies to infer the properties of magnetic flux on the
spatially-unresolved stellar surface.  The analog of the solar cycle in
late-type stars is typically seen through the modulation of chromospheric
radiative emissions, such as in the deep cores of the \ion{Ca}{2} resonance
lines, that are spatially associated with sites of emergent magnetic
fields. \citep{sku75,wilson78,bali95}. The amplitude modulation of this
magnetic flux is widely regarded as a property of non-linear dynamo processes
of which $\alpha-\omega$ kinematic dynamos are a particular class of mean-field
dynamo models \citep{tobias97}.

The high-quality photometric data from the space missions CoRoT 
\citep{Baglin-etal:09} %
and \textit{Kepler}
\citep{Koch-etal:10} %
have yielded new insight on the rotation and magnetic properties of solar-type
stars by providing rotation periods for thousands of main-sequence stars
\citep{McQuillan-etal:14,Nielsen-etal:13,Reinhold-etal:13,Buzasi-etal:16} %
as well as new photospheric proxies of magnetic activity based on the periodicity and
amplitude of the light-curve modulation 
\citep{He-etal:15}. %

In order to provide a broader parameter space for the advancement of stellar
and solar dynamo models, we further develop herein a method for the
measurement of active region area coverages on solar-type stars
\citep{giam85,ag95}.  In particular, we extend our previous work through the
results of simultaneously acquired, high-resolution spectroscopic observations
of the \ion{He}{1} triplet lines at 5876~\AA\ and 10830~\AA, respectively.  

\begin{figure*}
\begin{center}
\epsscale{.70}
\gridline{%
  \fig{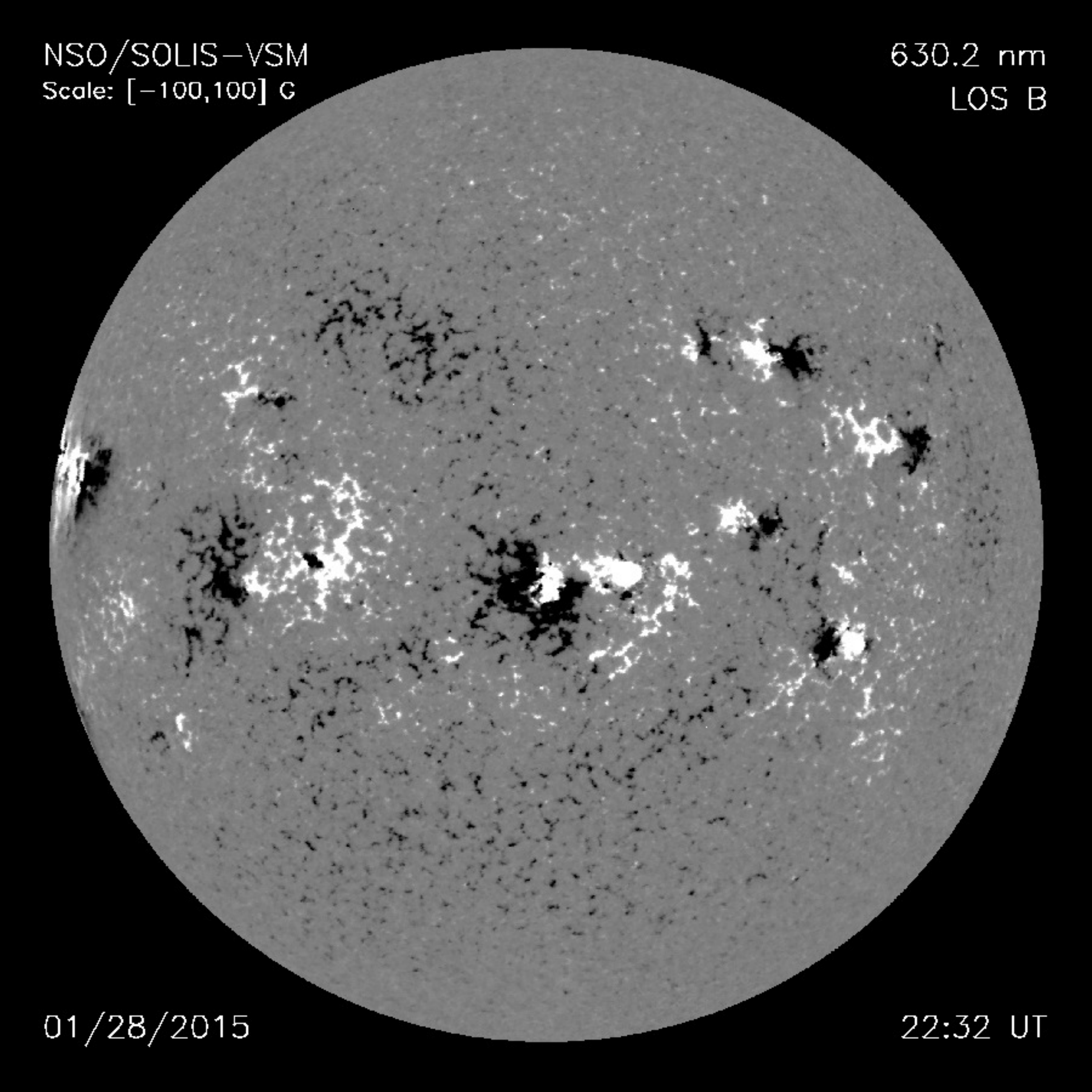}{0.24\textwidth}{}
  \fig{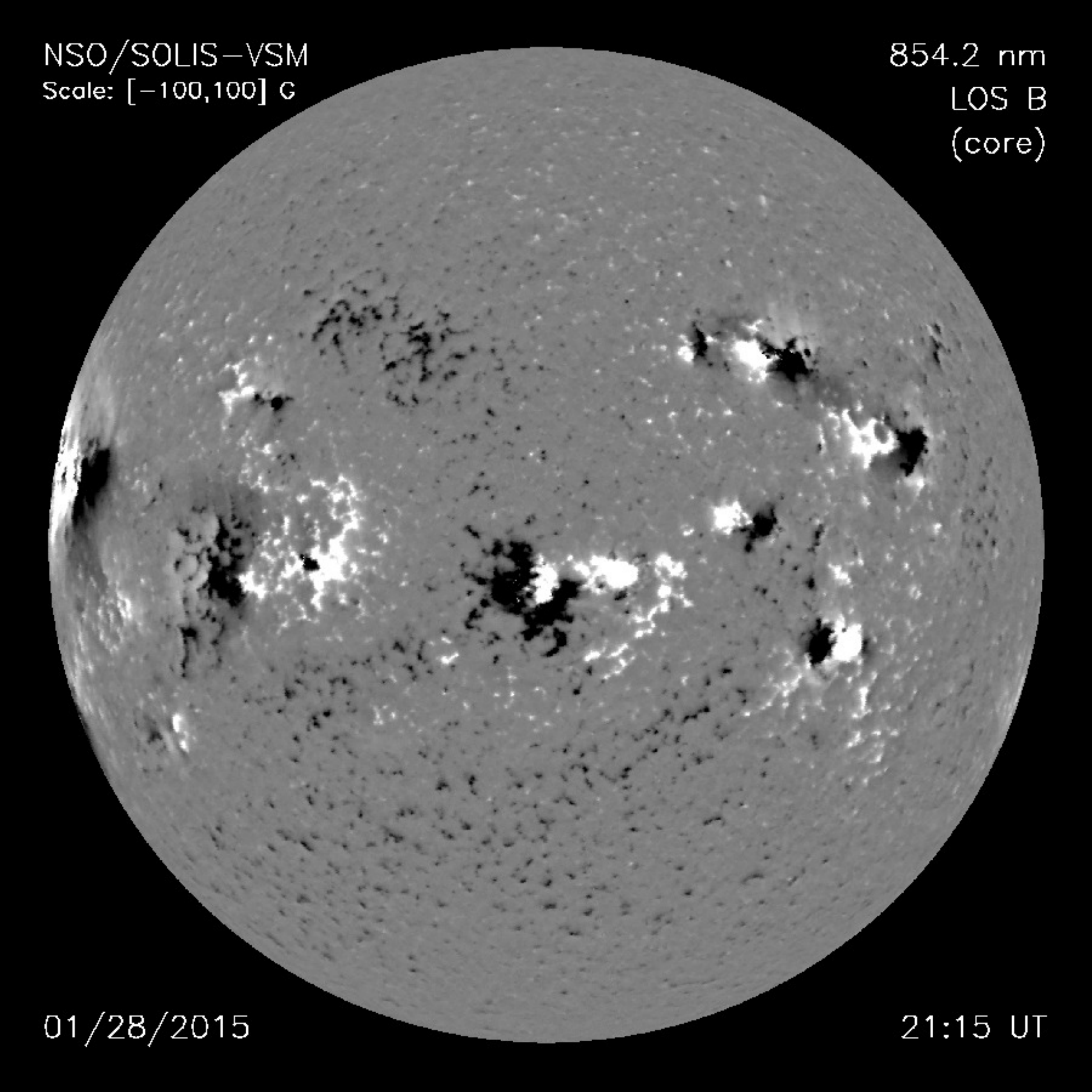}{0.24\textwidth}{}
  \fig{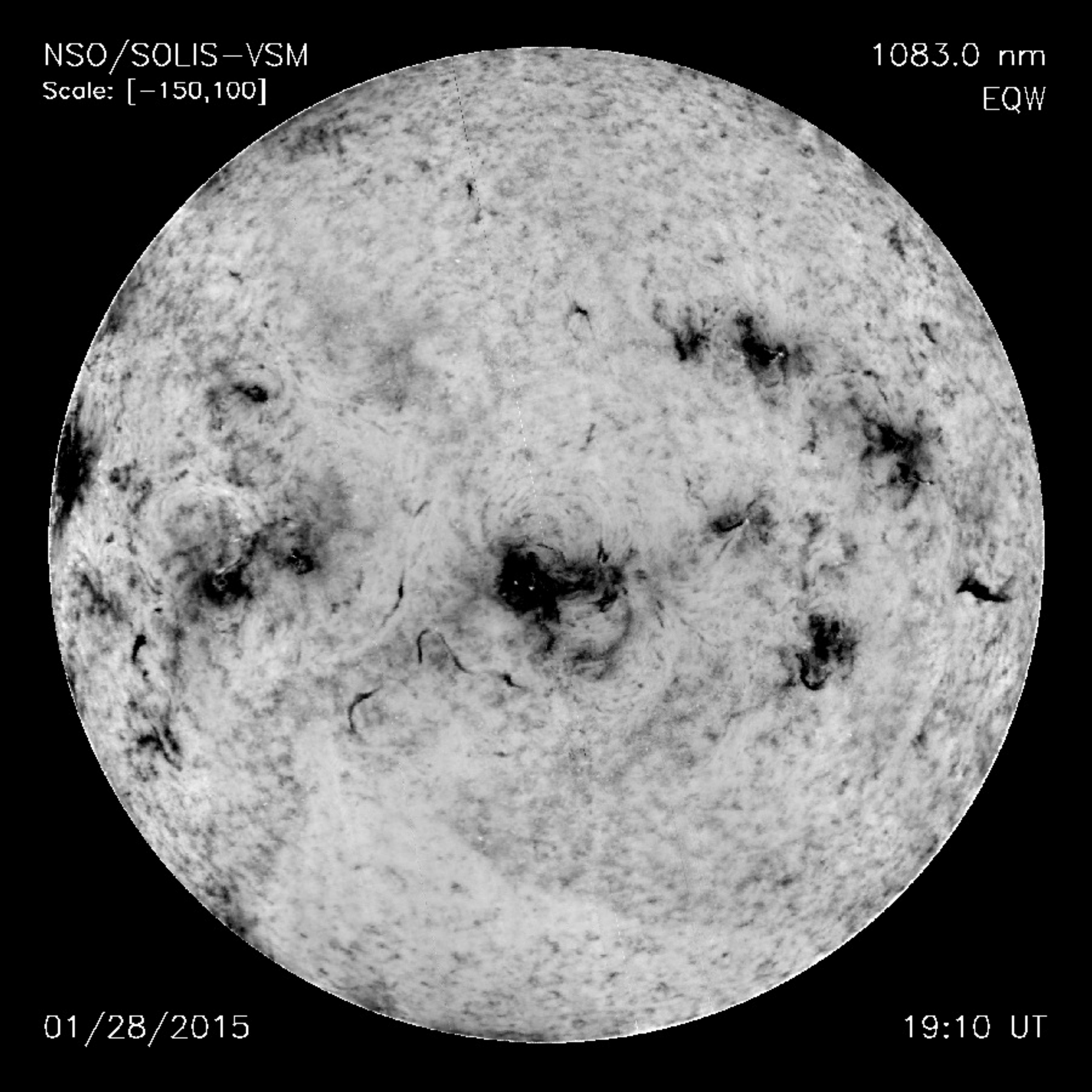}{0.24\textwidth}{}
  \fig{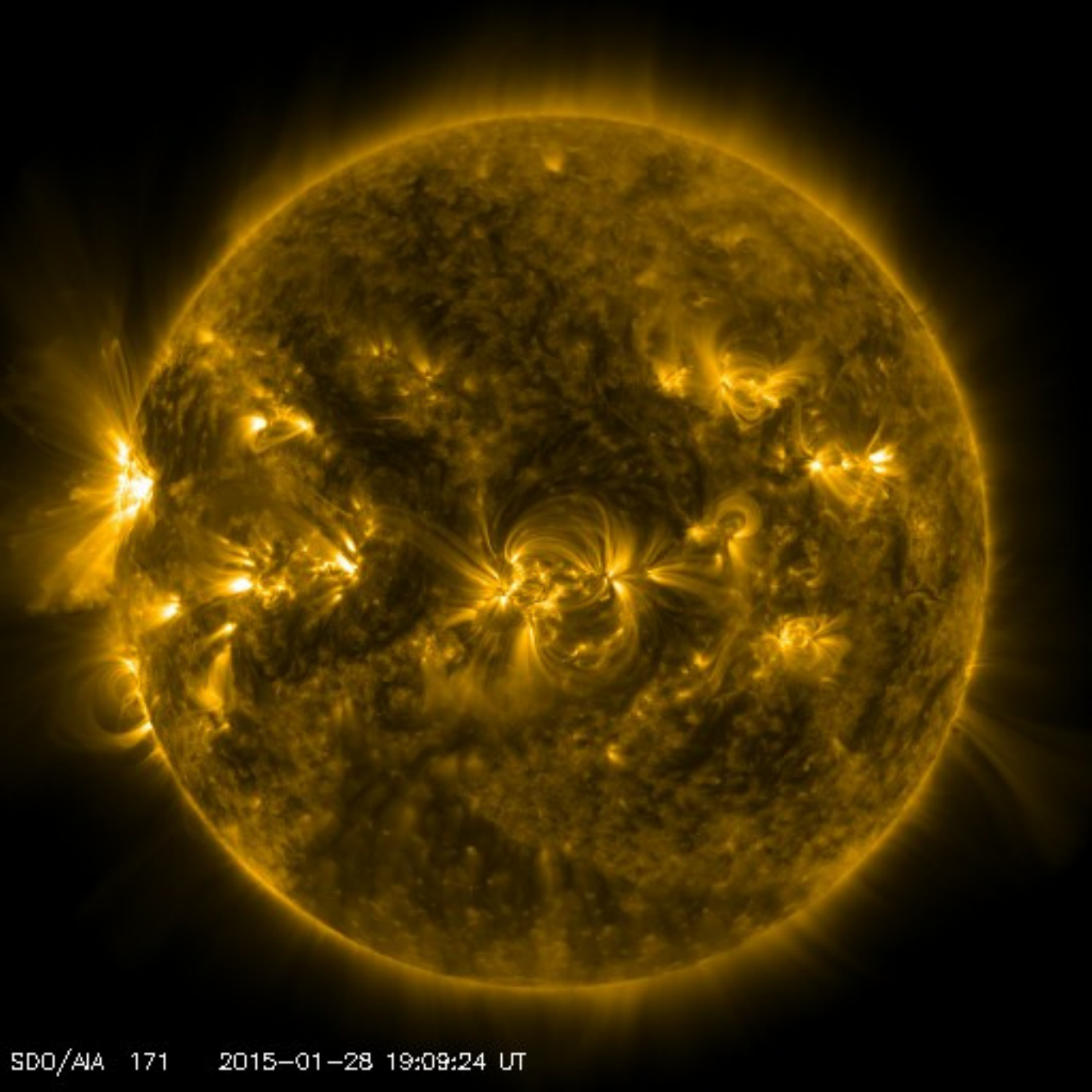}{0.24\textwidth}{(d)}
}
\caption{Full disk images of the Sun %
    on a date near the maximum of solar cycle 24.  From left to right:
    Longitudinal magnetograms in the photosphere (\ion{Fe}{1} 6302~\AA) and in
    the chromosphere (\ion{Ca}{2} 8542~\AA), a spectroheliogram in \ion{He}{1}
    10830~\AA, and an EUV image of the corona at 171~\AA.  The first three
    images are from the NSF SOLIS Vector Spectromagnetograph (VSM) while the
    coronal image is from the Atmospheric Image Assembly (AIA) instrument
    on-board the NASA Solar Dynamics Observatory (SDO) mission. %
  These images illustrate the spatial
  correlation between localized areas of relatively strong magnetic flux in
  both the photosphere and chromosphere, the presence of {\heir} absorption,
  and overlying coronal EUV emission.
  The {\heir} spectroheliogram appears more diffuse than the photospheric
  magnetogram, reflecting the greater height of formation in the chromosphere
  combined with field line spreading           and the effect of EUV
    back-illumination.  }
  \label{fig:full_sun}
\end{center}
\end{figure*}

Solar observations, such as those in Fig.~\ref{fig:full_sun}, demonstrate that
these lines are ideal tracers of magnetic regions outside of cool spots,
appearing in absorption in active regions and only weakly in quiet network
elements and the (non-magnetic) photosphere\footnote[3]{See \citet{jharvey75}
  for an analogous figure that includes more rare spectroheliograms in
  \ion{He}{1} \heiv\ ({\dthree}) obtained simultaneously with solar
  X-ray images from $Skylab$.}.  %

Thus, the measured absorption equivalent
width is proportional to active region area coverage, or filling factor (the two terms are
used interchangeably in this paper), at the height of
formation and the time of observation.  
A more precise estimate of the active region
filling factor can be obtained through examination of the joint response of
the \ion{He}{1} triplet lines to chromospheric heating combined with a
model-dependent calibration of the strengths that can be attained by these
features in plage-like regions on the stellar surface.
  Recall that plages are the chromospheric counterpart of faculae, which are
  localized bright regions in the solar photosphere associated with
  concentrations of magnetic fields and characterized by reduced opacity, thus allowing us to
  see the deeper, hotter (and, hence, brighter) walls of the facular area.
  The overlying plage is distinguished by relatively bright \ion{Ca}{2}~H \& K
  emission in full-disk spectroheliograms.  %

We discuss in Sec.~\ref{sec:theory} our approach to the calibration of
the joint response of the helium lines to atmospheric heating along with the
results of our model calculations.  The observations and their reduction are
presented in Secs.~\ref{sec:obs} and \ref{sec:reduction:spectra}.
A discussion of the inferred active
region filling factors is given in Sec.~\ref{sec:results}. In
Sec.~\ref{sec:summary} we present our conclusions and our anticipated
directions for further research.

\section{Model Approach}
\label{sec:theory}

  \citet[hereafter AG95]{ag95} developed a technique to address this problem
  by demonstrating that the non-linear response of the two main triplet
  \ion{He}{1} lines (at 10830~\AA\ and 5876~\AA) to chromospheric heating can
  be exploited to infer the fractional area coverage by active regions in
  solar-like stars.  Their approach was based on a two-component
  representation of the strength of activity diagnostics, where the observed
  equivalent width of a line, \Wobs, can be written in terms of the
  contributions from the quiescent atmosphere, \Wq, and the active
  (plage-like) atmosphere, $\Wa$, via a filling factor $f$, where
  \begin{equation}
    \Wobs = (1 - f)\,\Wq\:+\:f\,\Wa\; . \label{eq:Wobs}
  \end{equation} 
    Further details on the derivation of the above equation can be found in
    AG95.  We only note here that the difference in continuum intensities
    between the quiescent and active atmosphere is assumed to be negligible.
    At this level of approximation, this is a valid assumption in the Sun and, by analogy, in solar-type
    stars.

    In the two-component model described above, the quantities $\Wq$ and $\Wa$
    represent the \emph{average} values in the quiescent atmosphere and in
    active regions, respectively.  In both regions, the observed line
    strengths can of course vary on smaller scales.  In the quiet Sun, for
    example, chromospheric line strengths are typically distributed in a
    characteristic spatial pattern called the ``supergranular network''.  The
    main assumption of Eq.~\ref{eq:Wobs} is therefore that the average value of
    the strength of the activity diagnostics, $\Wa$, is the same in all active
    regions on the stellar surface independent of their area, i.e.\ small
    and large active regions are equally ``bright'' when imaged in the chosen
    activity diagnostics.  Fig.~\ref{fig:full_sun} shows that this assumption
    is plausible, but for most activity diagnostics this assumption can also be
    quantitatively verified \citep[e.g.][]{adz14}.

  In this approach, the filling factor, i.e.\ the fractional area
    covered by active regions, is therefore one of the fundamental parameters
    discriminating between stars with different observed activity levels.  In the
  favorable case of a negligible contribution from the quiescent atmosphere,
  the filling factor is simply $f=\Wobs/\Wa$.  But even in this
  case the filling factor cannot be determined unless the intrinsic line
  strength in stellar active regions is known. AG95 showed that this ambiguity
  can be resolved by observing two lines with different, non-linear
  dependences on the atmospheric activity.

  In order to apply the method described by AG95 to a pair of activity
  diagnostics such as the main \ion{He}{1} triplet lines, the dependence of
  the intrinsic strength of both lines on the active region heating needs to
  be computed, $\Wa(p)$, where $p$ is a parameter, or set of parameters
  characterizing the active regions in the formation layer of the diagnostics
  under consideration.  In AG95, the activity parameter is the mass loading,
  or column density, $m$, in g~cm$^{-2}$ at the top of the chromosphere or,
  equivalently, the increase of total chromospheric pressure relative to the
  quiescent state, $P/P_\mathrm{q}$.  A more complete set of activity
  parameters can be considered, as described below.

  In comparison with other diagnostics, the helium triplet lines are
  especially suitable for this approach since they respond to chromospheric
  heating, which is parameterized in model computations by higher chromospheric
  pressure in the line formation region, by increasing their absorption
  equivalent widths in a non-linear fashion (see AG95, their Fig.~2).  At
  sufficient densities, such as those that may occur in a flare, collisional
  control eventually overcomes scattering processes and 
  the triplet lines
  are driven into emission.  Thus, as atmospheric heating increases, all
  \ion{He}{1} triplet lines go deeper into absorption, reaching a maximum in
  their equivalent widths before going eventually into emission.  

\begin{figure}
  \epsscale{1.00}
  \includegraphics[angle=90,width=1.00\linewidth]{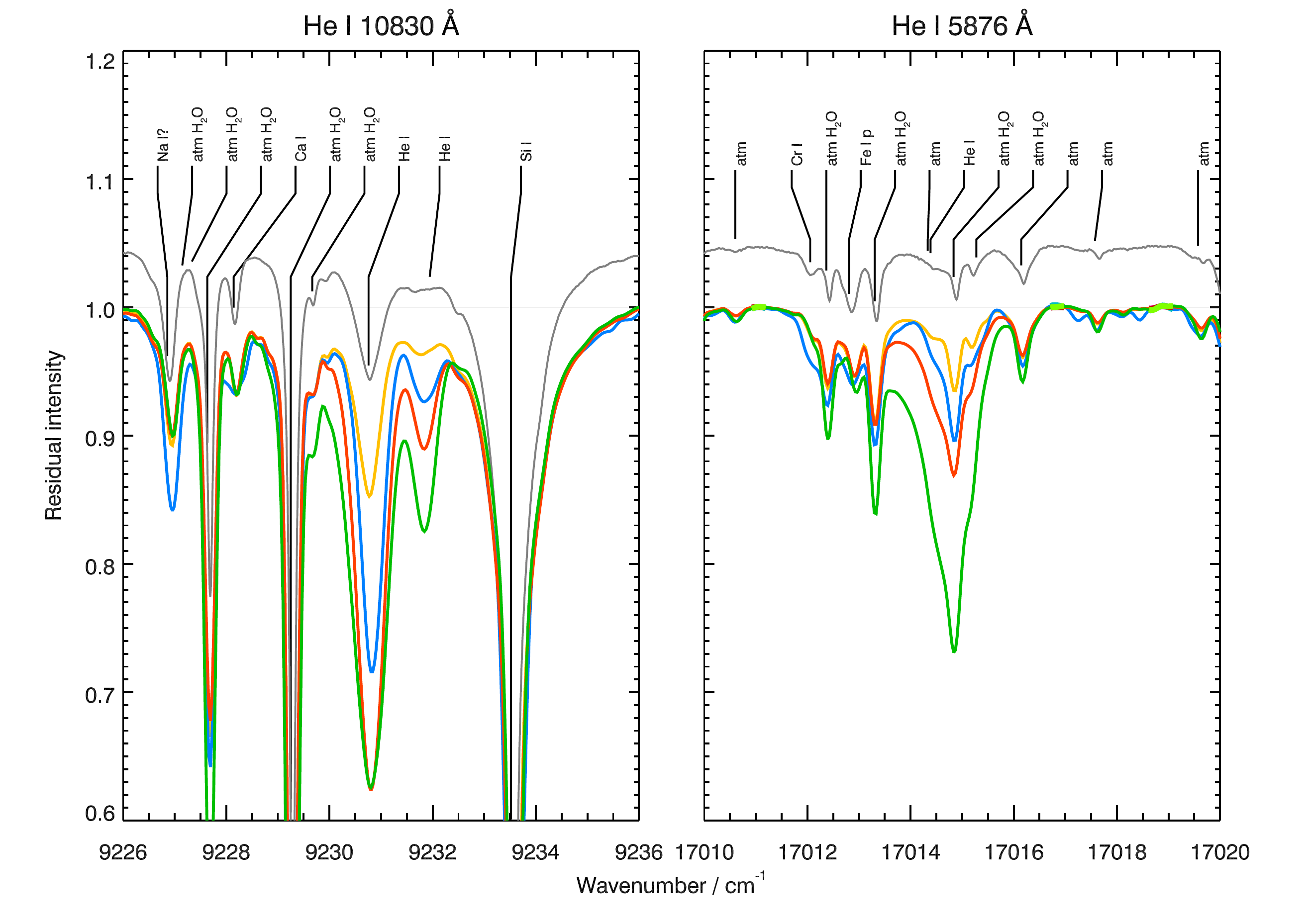}
  \caption{%
    Solar observations of the \ion{He}{1} triplet lines in various locations
    of an active region.  The data were obtained on 16 March 1995 with the
    Fourier Transform Spectrometer at the NSF McMath-Pierce Solar Telescope on
    Kitt Peak \citep{Brault79}.  The spectra refer to
    various locations of plages or plage-like areas of AR
    7854.  %
      On the left and right panels are marked, respectively, the \ion{He}{1}
      lines at 10830.3~\AA\ (air wavelength, corresponding to wavenumber
      9230.8 cm$^{-1}$, with its minor fine structure component at 10829.0~\AA, or at
      wavenumber 9231.9 cm$^{-1}$) and at 5875.7~\AA\ (corresponding to
      wavenumber 17014.5 cm$^{-1}$). In both panels the nearest solar and
      telluric lines are also marked, including the group of telluric water
      lines within the profile of the \ion{He}{1} \heiv\ feature.
    The reference solar flux atlas by \citet{kurucz84} is also displayed for
    comparison (grey) with an offset of 0.05 above the observed
      spectra.
  }
  \label{fig:solar_spectra}
\end{figure}

  This general behavior of the helium lines is in qualitative agreement with
  observations, as illustrated in Fig.~\ref{fig:solar_spectra}, where we see
  varying strengths in the triplet lines at different locations in a solar
  active region, presumably in response to
  different degrees of chromospheric heating.

  We note that this behavior is very much reminiscent of H$\alpha$ line
  formation in the chromospheres of M dwarf stars \citep{cm79,giam82}.
  Following the approach of \cite{giam85}, the existence of a maximum in
  intrinsic line strength in active regions, $\Wmax$, can be exploited to
  derive a lower limit to the filling factor:
  \begin{equation}
    f \ge \frac{\Wobs-\Wq}{\Wmax-\Wq}\; , \label{eq:fmin}
  \end{equation}
  or, if $\Wq\approx 0$, $f\ge\Wobs/\Wmax$.
  
  The key point of the method described in AG95 is however that each line
  attains its maximum equivalent width at different amounts of atmospheric heating.
  This leads to a strongly non-linear joint response of the line
  strengths.  Thus, simultaneous observations of two lines can in principle
  allow an unambiguous determination of the filling factor $f$.  This
  behavior is illustrated in Fig.~\ref{fig:solar_spectra}, where the main
  component of the \heir\ line seems to reach a level of ``saturation'' in its
  equivalent width while both the \dthree\ line and the minor component of
  the IR triplet line at 10829~\AA\ continue to increase in their strength.

\begin{figure}
  \begin{center}
    \epsscale{2.0}
    \includegraphics[angle=90,width=1.05\linewidth]{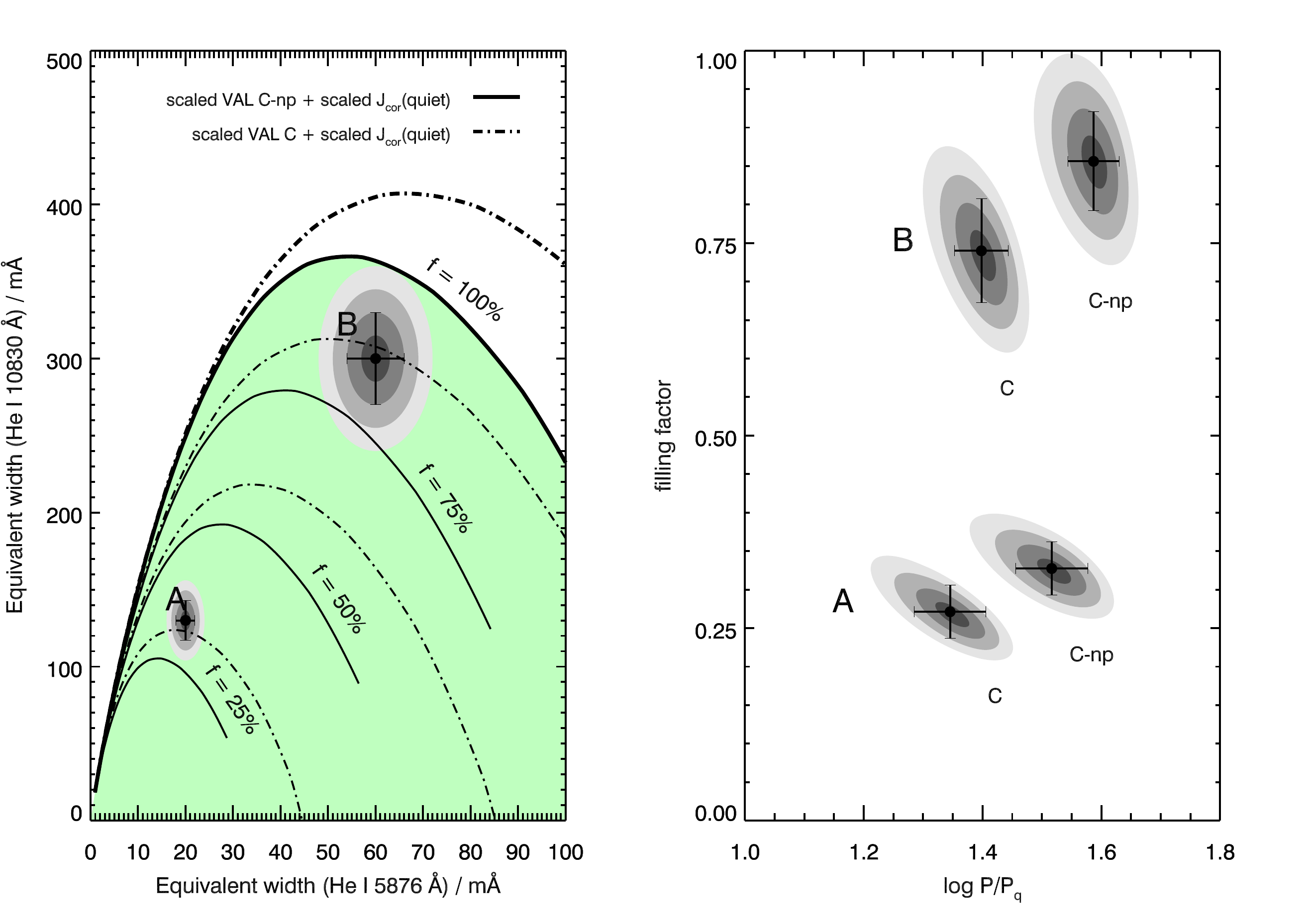}
    \caption{%
      Illustration of the method described in AG95 in the case of the
      joint response of the main \ion{He}{1} triplet lines.  %
      Two
        hypothetical observations, labeled A and B, are 
      also shown to illustrate the effect of
      uncertainties in the equivalent width measurements on the determination
      of the filling factor by this method.  
        The error bars and the corresponding bi-dimensional probability
        distribution in the left-hand panel correspond to a 10\% uncertainty
        in both equivalent widths.  The transformed distributions in the
        $(P/P_\mathrm{q},f)$ plane are shown in the right hand panel for the
        two series of theoretical models, labeled C and C-np, 
          that are described in Sec.~\ref{sec:theory:atmo}
        and shown in Fig.~\ref{fig:atmo_models}.  
        The dots and error
        bars represent, respectively, the mean and standard deviation of the
        transformed
        distributions of filling factor and pressure enhancements.
    }
    \label{fig:joint_response}
  \end{center}
\end{figure}

  The essence of the method is illustrated in Fig.~\ref{fig:joint_response},
  which displays theoretical diagrams as calculated by AG95 of the joint
  response of the triplet lines in equivalent width to chromospheric heating
  (dot-dashed lines), together with the set of calculations adopted here and
  described in Sec.~\ref{sec:theory:atmo} below.
  The locus $f=1$ defines a region (highlighted in solid color in the figure for the
  reference calculations) where all measurements should fall; already in AG95
  it was shown that observations of solar-like stars do indeed fall in this
  allowed region.  We also note that to infer the filling factor it is not
  necessary to have a detailed knowledge of the specific activity state of the
  stellar plage-like regions; only the joint dependence of the two spectral
  diagnostics.  Nevertheless, the values of the activity parameter $p$ 
  that best match the observations can still be derived together with $f$ by
  inverting Eq.~\ref{eq:Wobs} for the selected activity diagnostic pair.

    The effect of measurement errors is also shown in
    Fig.~\ref{fig:joint_response}.  Two hypothetical joint measurements of
    \heiv\ and \heir\ with a 10\% ($1\sigma$) uncertainty are shown in the
    left-hand panel of that figure.  The corresponding bi-dimensional
    probability distributions are shown as filled gray contours.  In the
    right-hand panel, the probability distribution transformed by the inversion
    of Eq.~\ref{eq:Wobs} for both lines is shown in the
    ($P/P_\mathrm{q}$,$f$) plane, for both sets of theoretical calculations of
    $\Wa(P/P_\mathrm{q})$ we have considered and that are discussed in
    Sec.~\ref{sec:theory:atmo} below.  The mean value and the standard
    deviation of the transformed distributions are shown as error bars.  In
    particular, the mean and the standard deviation of the transformed
    probability distribution for the filling factor is the value and its error
    we will attach to the actual measurements described in the remainder of
    this work.

  In addition to the general properties of their joint response to
  chromospheric heating, the helium triplet lines exhibit several desirable
  features:
  \begin{itemize}
  \item they are purely chromospheric lines: the photospheric contribution to
    these lines is negligible in solar-like stars;
  \item their strength in the quiescent chromosphere is small: the observed
    lines in spectra of solar-like stars arise almost entirely in
    active 
    regions: $\Wq(\dthree) \approx 0$ and $\Wq(\heir) \approx 40$~m\AA\ 
    (the latter 
    value is inferred from full-disk measurements during the minima of solar
    activity: \citealt{harvey-livingston:94,livingston-etal:10});
  \item they both belong to the same atom: therefore the effect of the elemental
    abundance is largely factored out;
  \item the transitions giving rise to the two lines share one atomic level
    ($1s2p$ $^3\!P$) in so-called orthohelium: thus their differential
    behavior is relatively insensitive to the details of interactions with other
    atomic levels;
  \item they form essentially in the same zone of the chromosphere, regardless
    of the details of the formation mechanism \citep{agj95,aj97}.  Hence, they
    probe exactly the same regions of the stellar atmosphere.
  \end{itemize}

  \subsection{Reference calculations}
  \label{sec:theory:atmo}

  In AG95 the atmospheric activity level is parameterized by the column
  density, $m$, in g~cm$^{-2}$ at the top of the chromosphere.  This
  formulation has the advantage that the total chromospheric pressure, $P$, is
  simply given by $P=gm$, where $g$ is the stellar surface gravity.  Implicit
  in this relation is the assumption that the chromosphere is thin with
  respect to the stellar radius.  In a parallel study, \citet[hereafter
  AJ97]{aj97} carried out a more extensive analysis of the parameters
  determining the formation of the helium spectrum in the Sun.  

  The reference quiescent model adopted in both AG95 and AJ97 is the VAL~C model of
  the quiet Sun \citep{val81}.  In AJ97, two modified versions of the model,
  termed VAL~C-np and VAL~C-nt, were also considered, which differ from the
  VAL~C model only in the thickness of the transition region, $\Delta
  h(\mathrm{TR})$.  The C-nt series, i.e.\ the series starting from the
  VAL~C-nt model was used in AJ97 only to discuss some specific radiative
  transfer aspects of the line formation; we will not consider this series of
  models here.

  Both AG95 and AJ97 included in their
    analysis the effect on the helium ionization balance of
  coronal EUV back-illumination integrated in the range $\lambda <$ 500~\AA\
  ($J_\mathrm{cor}$).  In AG95 the EUV
  back-illumination was suitably scaled to account for increased coronal
  emission in active regions.  The same scaling was applied by
  \citet[hereafter A94]{a94} also for the C-np series of atmospheric models.
  Hence, the pair $\{m/m_\mathrm{q},\Delta h(\mathrm{TR})\}$
    constitutes the fundamental set of parameters, $p$, determining 
    both the structure of the atmosphere and the coronal back-illumination,
    that, in turn, 
  are used to compute the theoretical equivalent widths of the
  \ion{He}{1} triplet lines to be used in Eq.~\ref{eq:Wobs}, $\Wa(p)$.

\begin{figure*}
  \begin{center}
    \epsscale{2.0}
    \includegraphics[angle=90,width=1.05\linewidth]{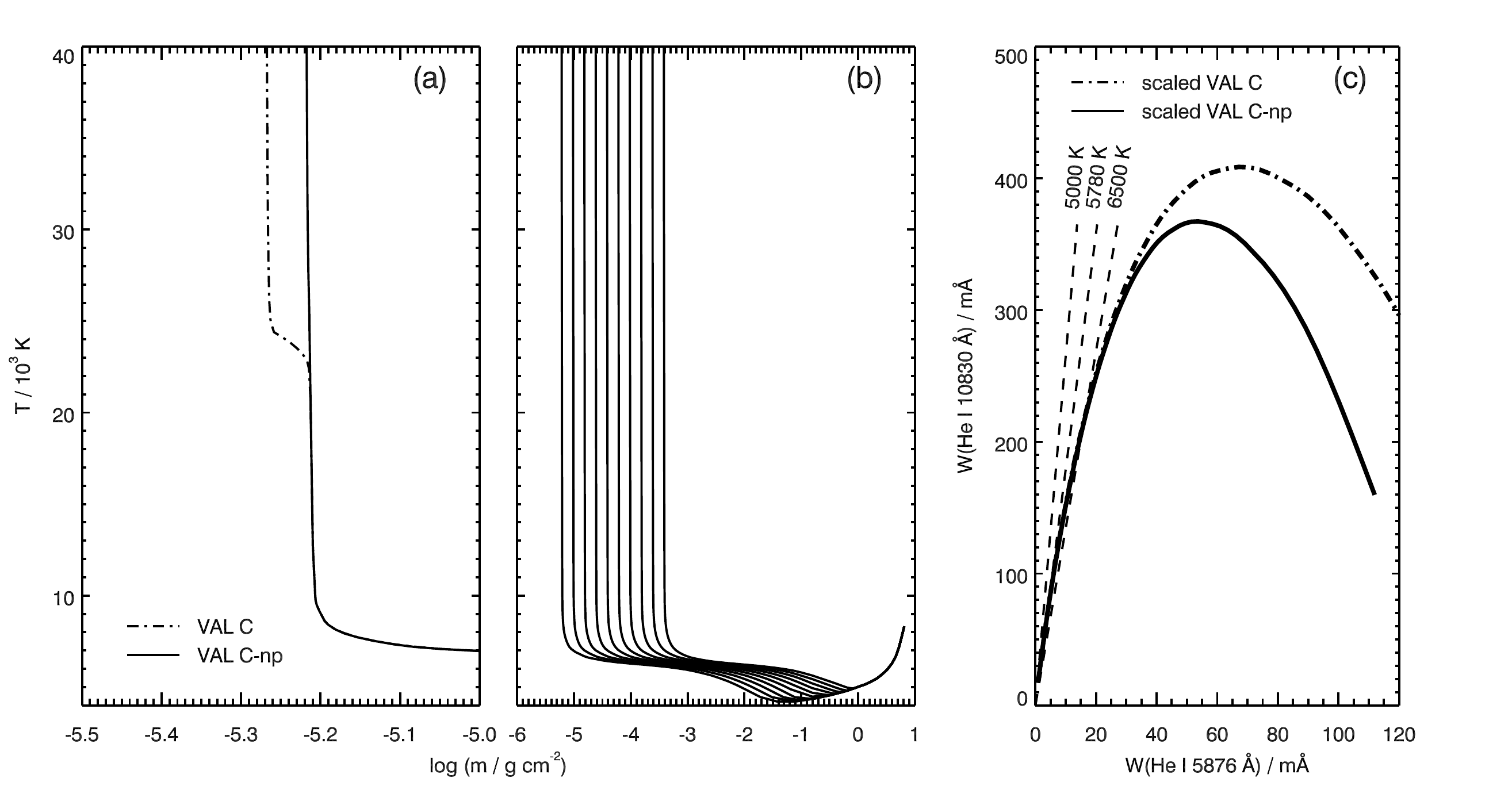}
    \caption{Model calculations.  %
      (a) %
      Two examples of quiescent model chromospheres from AJ97 in temperature
      vs. mass column density, $m$. 
      (b) %
      The series of models derived from the VAL~C-np quiescent model by increasing
      the mass loading at the top of the atmosphere.
      (c) %
      The computed theoretical diagrams of $W(\heir)$ vs $W(\dthree)$ 
      at $f = 1$,  
      for a star with \Teff= 5800 K.
      The optically thin limit of the joint response of the two lines is
      also shown by dashed lines, as functions of stellar effective temperature.
      See Sec.~\ref{sec:theory:atmo} for a detailed description of the models.
    }
    \label{fig:atmo_models}
  \end{center}
\end{figure*}

  In AJ97 it was shown that the C-np series, i.e.\ the series of atmospheres
  with a reduced temperature plateau at $\approx 2\times10^4$~K, 
  better matches
     the observed properties of the solar \ion{He}{1} spectrum
  than the C and C-nt series, from the extreme UV (EUV) to the IR.  That
  finding is consistent with the structure of solar plages derived from
  semiempirical models \citep[e.g.:][]{fal93,fea06}. 
  We therefore adopt the C-np series as our reference set of models,
  considering however the scaling of EUV back-illumination as computed by A94
  and AG95.  We nevertheless also take into consideration the C series of
  models, i.e.\ the series based on the VAL~C model, as in AG95, for
  comparison.  Figure~\ref{fig:atmo_models} shows the series of models
  computed in A94 (panels a and b) which were then employed by AG95, AJ97, and
  in the present work, together with the corresponding joint response of the
  triplet lines as functions of the parameter $P/P_\mathrm{q}$ (panel c).

    The effect of the choice of the series of models on the determination of
    the filling factors is illustrated in Fig.~\ref{fig:joint_response}.  For
    the two examples shown, the differences introduced by the different
    theoretical calculations of $\Wa(P/P_\mathrm{q})$ are larger than the
    uncertainties due to measurements errors, if the latter are of the order
    of 10\% or less.

    We note, however, that the maximum equivalent width attained by the
    \ion{He}{1} \heir\ line is very similar in the two series of models:
    $\approx 370$~m\AA\ for the C-np series, and $\approx 410$~m\AA\ for the C
    series.  
      From an observational point of view, \cite{SanzForcada-Dupree:08} noted
      that data for cool dwarfs and subdwarfs tend to be below those
      theoretical limits, with very few exceptions in very active binaries.  A
      similar result is obtained by inspecting the data of
      \cite{Zarro-Zirin:86}.
 
    Regarding the quiescent value for the \ion{He}{1} \heir\ line adopted here
    ($\Wq=40$~m\AA), we note that data published in \cite{Zarro-Zirin:86} for
    low activity stars for which some \ion{He}{1} \heir\ absorption could be
    detected tend to cluster around $\sim 50$~m\AA\ (see their figures 1a and
    1b).  Values reported by \cite{takeda-takadahidai:11} have a median of
    $35$~m\AA, while values reported by \cite{GSmith-etal:12} are around $\sim
    30$~m\AA.  These last two papers present mostly data for low-metallicity,
    low-gravity stars whose atmospheres could significantly differ from those
    of solar-like stars as far as the relevant regions contributing to the
    formation of the \ion{He}{1} optical lines are concerned (photosphere,
    transition region, corona.)  Furthermore, the relatively modest variations
    in the minimum detected \heir\ equivalent width have a little effect on
    the values obtained with Eq.~\ref{eq:fmin}, since $\Wmax$ is about an
    order of magnitude larger than $\Wq$.

  In conclusion, the lower limits of the filling factors derived from Eq.~\ref{eq:fmin}
  are practically insensitive to the details of the adopted models.  On the
  other hand, the \dthree\ line never attains its maximum in the grid of
  models considered by AG95 and AJ97 and therefore a similar approach
    based on \dthree\ measurements alone is not feasible in solar-like stars.

    Concerning the dependence of the joint response of the \ion{He}{1} triplet
    lines on stellar effective temperature, the calculations of AG95 for an
    F star with $\Teff=6500$~K show a slightly lower slope of the
    initial linear part of the curve compared to the case of the Sun,
    considered as a typical G-type star.  This behavior can be understood
    given that at low activity levels the line formation is dominated by
    scattering of photospheric radiation (see discussion in A94 and AJ97).  
    The relevant photoexciting radiation determining the slope of
      the linear part of the joint response of the \heir\ and \heiv\
      lines is the photospheric radiation field at 10830~\AA. %
    Following the same
    argument, we expect calculations for chromospheres illuminated by the
    photosphere of a K-type star to show a slightly steeper joint response of
    the two triplet lines at low activity levels. 
      The effect of the photospheric radiation on the joint response of the
      two \ion{He}{1} triplet lines is shown in Fig.~\ref{fig:atmo_models}
      for the optically thin case.%

  Finally, we note that a number of theoretical and observational studies on
  the formation of the helium spectrum in the Sun have appeared since 
  A94, AG95, and AJ97 \citep[e.g.:][]{%
    macpherson-jordan:99,%
    andretta-etal:00,%
    smith-jordan:02,%
    smith:03,%
    andretta-etal:03,%
    pietarila-judge:04,%
    judge-pietarila:04,%
    mauas-etal:05,%
    andretta-etal:08}. %
  Most of those studies, however, were focused on the formation of the EUV
  lines and continua, while the mechanism responsible for the formation of the
  optical subordinate lines has attracted comparatively less
  attention, with some recent exceptions such as \cite{Leenaarts-etal:16}.  
  In
  any case, we remark that in this investigation we are merely utilizing those earlier
  calculations, and that updating the models is beyond the scope of this 
  paper.

\section{Observations}
\label{sec:obs}

Given the potential effects of variability due to magnetic activity on the
strengths of the triplet features, combined with a method based on
observations of the joint behavior of these diagnostics, our observational
approach was to obtain 
spectra of the \dthree\ line in the visible
and the Near-Infrared (NIR) \heir\ line on the same night, respectively. %
  Obtaining simultaneous spectra of the two lines is a challenge even in the
  case of the Sun, but it is nevertheless feasible, as the spectra of
  Fig.~\ref{fig:solar_spectra} demonstrate.  In addition to those FTS
  spectra, to our knowledge only \cite{muglach-schmidt:01} have been
  successful in obtaining simultaneous observations in the two lines.  On the
  other hand, we could not find analogous observations of solar-like stars in
  the literature, although both the \ion{He}{1} \heiv\ and the \heir\ lines
  have been extensively studied in the context of stellar activity, as we
  briefly recap in the following.

  Guided by \ion{He}{1} \dthree\ spectra obtained for solar plages 
  \citep{Landman81},
  extensive stellar observations of \dthree\ as an activity diagnostic
  utilizing digital detectors with peak sensitivities in the visible soon
  followed.  \cite{lambert-obrien:83}
  reported the detection of rotational
  modulation of \dthree\ in selected main sequence stars.  
  \cite{wolff-heasley:84}
  conducted a survey of \dthree\ in a sample of G and K stars followed
  by a survey focused on main-sequence stars 
  \citep{wolff-etal:85}.
  These investigations were soon followed by focused studies addressing
  specific questions.  Examples include the determination of the effective
  temperature on the main sequence corresponding to the onset of chromospheric
  activity associated with outer envelope convection 
  \citep{wolff-etal:86,wolff-heasley:87,garcialopez-etal:93};
  the
  correlation of \dthree\ absorption strength with rotation as well as its
  empirical relationship with other diagnostics of magnetic-field related
  activity such as X-ray emission
  \citep{saar-etal:97};
  and, evidence for cycle-like variability in \dthree\ seen in multi-year
  stellar programs involving high precision radial velocity monitoring 
  \citep{santos-etal:10}.
  While a stronger feature,                  for a long time  studies of the
  \ion{He}{1} \heir\ line in late-type, dwarf stars                  have been
    more  limited due to the lack of sensitivity of available detectors in
  this spectral region though probes of chromospheric structure based on
  \heir\ have been carried out in recent years with large-aperture telescopes
  \citep[e.g.:][]{takeda-takadahidai:11,smith-etal:12}.
  To our knowledge, there are no near-simultaneous observations of both {\dthree} and
  \heir\ %
    in stars displaying solar-like activity, while such data exist for T~Tauri
    stars \citep[e.g.:][]{Dupree-etal:12}.
The primary challenges in the utilization of the helium triplet lines are that
(a) they are intrinsically weak and (b) they are blended with terrestrial
water lines.  These issues are best addressed with very high quality spectra
(in terms of S/N ratio and resolution) acquired at very dry sites to mitigate
the effects of terrestrial water vapor contamination.  Even when these
requirements are met, the presence of blends with nearby atomic lines in the
stellar spectrum due to rotational smearing 
(which is typically larger in more active late-type stars)
introduces an
additional source of error in the estimates of equivalent width.
  In addition to the difficulties of observing each line individually, their wide
  wavelength separation adds further challenges in obtaining simultaneous
  spectra with the same spectrograph.  

  In view of these considerations, we utilized the 8.2-m Very Large Telescope
  (VLT) and the  
  CRyogenic high-resolution InfraRed Echelle Spectrograph 
  (CRIRES) at the
  European Southern Observatory (ESO) at Cerro Paranal to obtain 
    the NIR \heir\ spectra 
    on the night of UT 2011 December 6 -- 7.
    The \dthree\ spectroscopic observations were carried out on the same night
    using the Fiber Extended-range Optical Spectrograph (FEROS), mounted at
    the 2.2m Max-Planck Gesellschaft/European Southern Observatory (MPG/ESO)
    telescope at La Silla (Chile), during MPG guaranteed time.

\subsection{Target selection}
\label{sec:obs:targets}

The principal selection criteria for the stellar sample included visually
bright (V $<$ 7) F, G, and K dwarfs that are detected X-ray sources in the
ROSAT All-Sky Bright Source Catalogue \citep{voges99} or listed in the
Gliese-Jahreiss Catalogue of Nearby Stars \citep{gj95}. The application of the
large-aperture VLT to bright objects served the dual objectives of efficiently
obtaining spectra of the highest quality for a large number of targets in a
single allocated night.  The target selection criteria are clearly biased
toward active stars since it was our intention to obtain spectra with
detectable helium triplet lines in order to further develop our analysis
methods as opposed to carrying out a survey at this time according to some
completeness criteria.

  The time differences between FEROS and CRIRES spectra of the same target are
  below 30 minutes, with the exception of HD~17051 and HD~33262 for which the
  time difference is about 1 hour.
A journal of the observations is presented
in Table~\ref{tab:obslog}. 
Note that the
CRIRES observations at \heir\ terminated earlier due to the onset of
adverse weather at the Cerro Paranal site, resulting in fewer targets observed
than at the La Silla site where the \dthree\ spectra were obtained with FEROS and
the ESO 2.2-m telescope. 
The details of the observations are given below.

\begin{deluxetable*}{llrllrlc}
\tablenum{1}
\tabletypesize{\scriptsize}
\tablewidth{0pt}
\tablecaption{Journal of Observations. The UT date for all start times is 2011 December 7. Label `std' in the notes indicate telluric standards. Table~\ref{tab:obslog} is published in its entirety in the machine readable format. A portion is shown here for guidance regarding its form and content.}
\tablehead{
\colhead{HD} & \colhead{Sp. type\tablenotemark{a}} & \colhead{\BV\tablenotemark{a}} & \multicolumn{2}{c}{UT start time} & \colhead{Nod. pos.} & \colhead{Notes} \\
\colhead{} & \colhead{} & \colhead{} & \colhead{FEROS} & \colhead{CRIRES} & \colhead{} & \colhead{}
}
\startdata
HD 49933   & F3V                  &       0.36 &   06:22:47     &              &           &            \\
           &                      &            &   06:25:23     &              &           &            \\
           &                      &            &   06:27:59     &              &           &            \\
HD 29992   & F3IV                 &       0.37 &   03:19:05     & 03:40:28     & A         &            \\
           &                      &            &   03:20:41     & 03:42:28     & B         &            \\
           &                      &            &   03:22:19     &              &           &            \\
HD 37495   & F5V                  &       0.46 &   05:07:55     & 05:10:26     & A         & No AO      \\
           &                      &            &   05:10:01     & 05:15:42     & B         & No AO      \\
           &                      &            &   05:12:07     &              &           &            \\
HD 27861   & A1V                  &       0.08 &   06:00:11     & 00:55:58     & A         & std        \\
           &                      &            &   06:02:23     & 00:58:43     & B         & std        \\
           &                      &            &   06:04:35     &              &           & std        \\
HD 18331   & A1V                  &       0.09 &   02:55:09     &              &           & std        \\
           &                      &            &   02:57:31     &              &           & std        \\
           &                      &            &   02:59:53     &              &           & std        \\
\enddata
\tablenotetext{a}{From the Bright Star Catalog \citep{bsc91} or the Simbad data base}
\label{tab:obslog}
\end{deluxetable*}

\subsection{FEROS observations and data reduction}
\label{sec:obs:feros}
  FEROS is a bench-mounted, thermally controlled instrument, fed by two fibers
  providing simultaneous spectra of either the object and wavelength
  calibration or the object and sky.  It is designed to achieve high
  resolution ($R$=48,000), high efficiency ($\approx$ 20\%), and to provide an
  almost complete spectral coverage from ~3500 to 9200~\AA\ spread over 39 echelle
  orders \citep{kaufer-etal:00}.
  The entrance aperture of the fiber has a projected diameter on the sky of
  2.0\arcsec.  As the cross-disperser is a prism, the spectral orders are
  strongly curved on the CCD.  The detector is an EEV 2k4k CCD.

  A total of 134 FEROS spectra of our targets
  (including telluric standards) 
  were acquired
  using the object-sky mode to avoid contamination by Th-Ar lines, as for our
  purposes it was preferable to analyze clean spectra rather than attaining the
  highest radial velocity precision.   The integration times of individual exposures ranged between 12
  and 420~s to obtain a signal-to-noise ratio (SNR) greater than
  200 for stars brighter than V=7.

  The data were reduced using a modified version of the FEROS Data Reduction
  System (DRS) pipeline, implemented within ESO-MIDAS\footnote[4]{Munich Image
    Data Analysis System} (vers. 09SEPpl1.2) under context FEROS, which yields
  a wavelength-calibrated, normalized, one-dimensional spectrum.
    The details of the reduction steps are given by \cite{schisano-etal:09}.

  For each target, a triplet of consecutive spectra was obtained, with the
  exception of HD 88697.  Pixels with unusually high values were identified by
  comparing the three spectra, and then flagged as missing.  The triplet of
  spectra was then averaged to obtain a single spectrum with greater SNR.
  The estimated SNR in reduced spectra reach values of the order of
    1000, with a mean value around 650.
\subsection{CRIRES observations and data reduction}
\label{sec:obs:crires}
  The \ion{He}{1} \heir\ spectroscopic observations were carried out in
  visitor mode on the same night as the FEROS \dthree\ observations, using 
  CRIRES %
    \citep{crires04,crires06}, %
  mounted at
  Unit Telescope 1 (Antu) of the VLT array at Cerro Paranal.  The entrance
  slit width was set to 0.2\arcsec\ to attain a nominal resolving power of $R
  = 10^5$.
  The CRIRES science spectra are recorded on an array of four 1024$\times$512
  Aladdin III detectors.  The grating position (\#52) was chosen so that the
  \ion{He}{1} \heir\ line was recorded on detector \#3.  We verified that
  spectra on that detector were free of significant ghost features.

  Each star was observed at two nodding offset positions along the slit, A and
  B, with jitter.  The total exposure times of each science frame (without overhead)
  range from 2 to 10 s to obtain a SNR exceeding
  200 for the target stars.  Almost all the science frames
  were obtained with the Adaptive Optics (AO) system on to optimize the
  SNR; only the last few spectra of the observing run
  were obtained with the AO off,
  because of the increasingly deteriorating seeing
  due to the onset of adverse weather which eventually caused early
  termination of the run.

  Data reduction of each CRIRES frame was performed using the ESOREX pipeline for CRIRES.
  Science frames and flat-field frames were corrected for non-linearity and 1D
  spectra were extracted from the combined flat-fielded
  frames 
  with an optimal extraction algorithm.  The wavelength solution is based 
  on the Th-Ar calibration frames provided by ESO.  
    The wavelength solution was then refined in the vicinity of the \heir\
    line by matching the average positions of the strongest H$_2$O lines observed in
    the range 10772--10868~\AA\ with the wavelengths given by
    \cite{breckinridge-hall:73}.
  The estimated SNR in reduced spectra are in the range 200 -- 700.

\section{Spectral analysis} 
\label{sec:reduction:spectra}

\begin{figure*}
  \begin{center}
    \includegraphics[angle=90,width=1.05\linewidth]{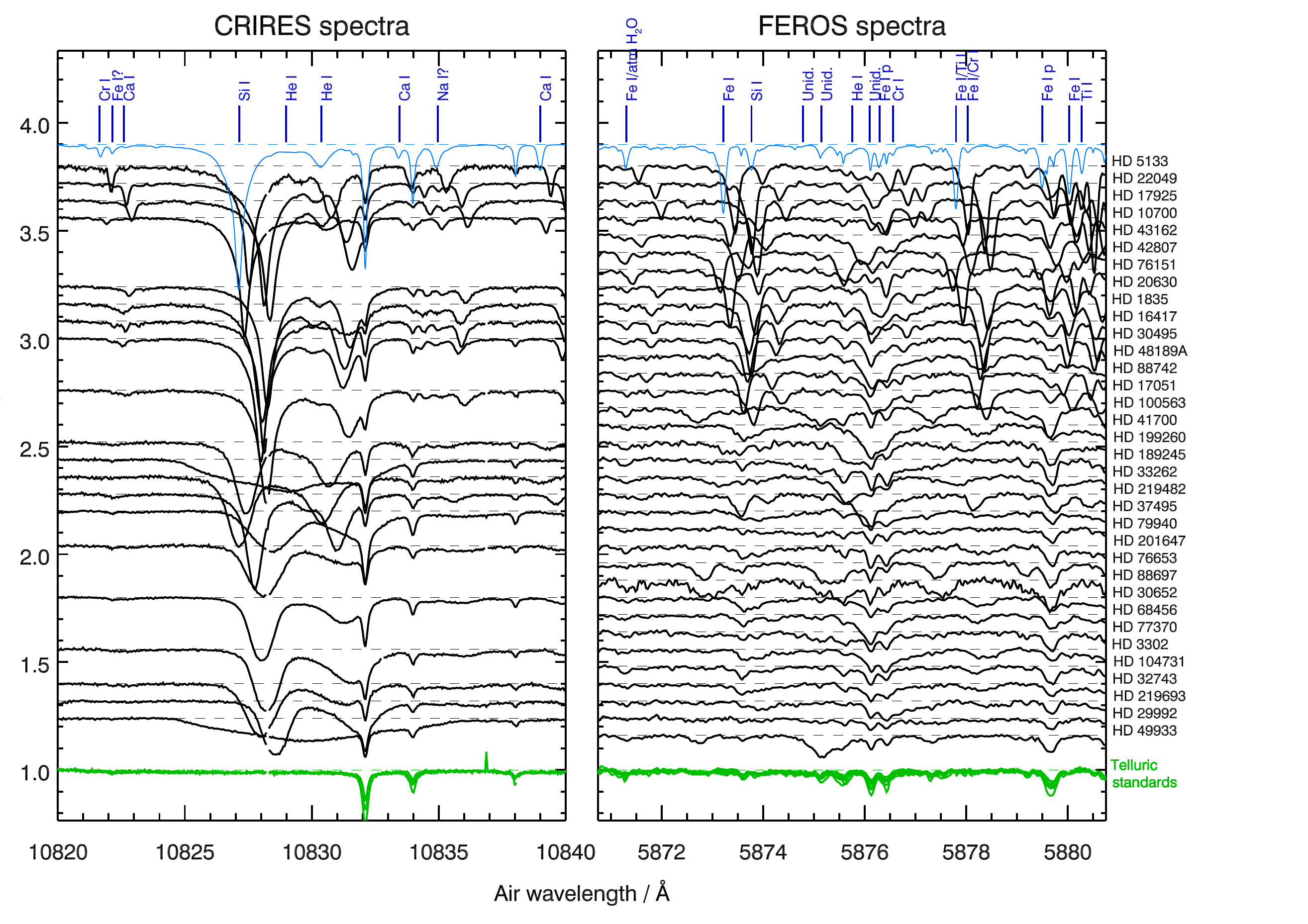}
    \caption{%
        Target spectra around the wavelength regions of interest.  
        Normalized fluxes are offset by a constant value.
        Spectra of telluric standards are shown at the bottom of the plot,
        highlighting the main telluric H$_2$O absorption lines in the range.
        The reference solar flux atlas by \citet{kurucz84} is also displayed
        for comparison as in Fig.~\ref{fig:solar_spectra} with identification
        of the main solar lines in the range, from \cite{swensson-etal:70} and
        \cite{moore-etal:66}. 
    }
    \label{fig:spectra}
  \end{center}
\end{figure*}

\begin{figure*}
  \begin{center}
    \epsscale{2.0}
    \includegraphics[angle=90,width=1.05\linewidth]{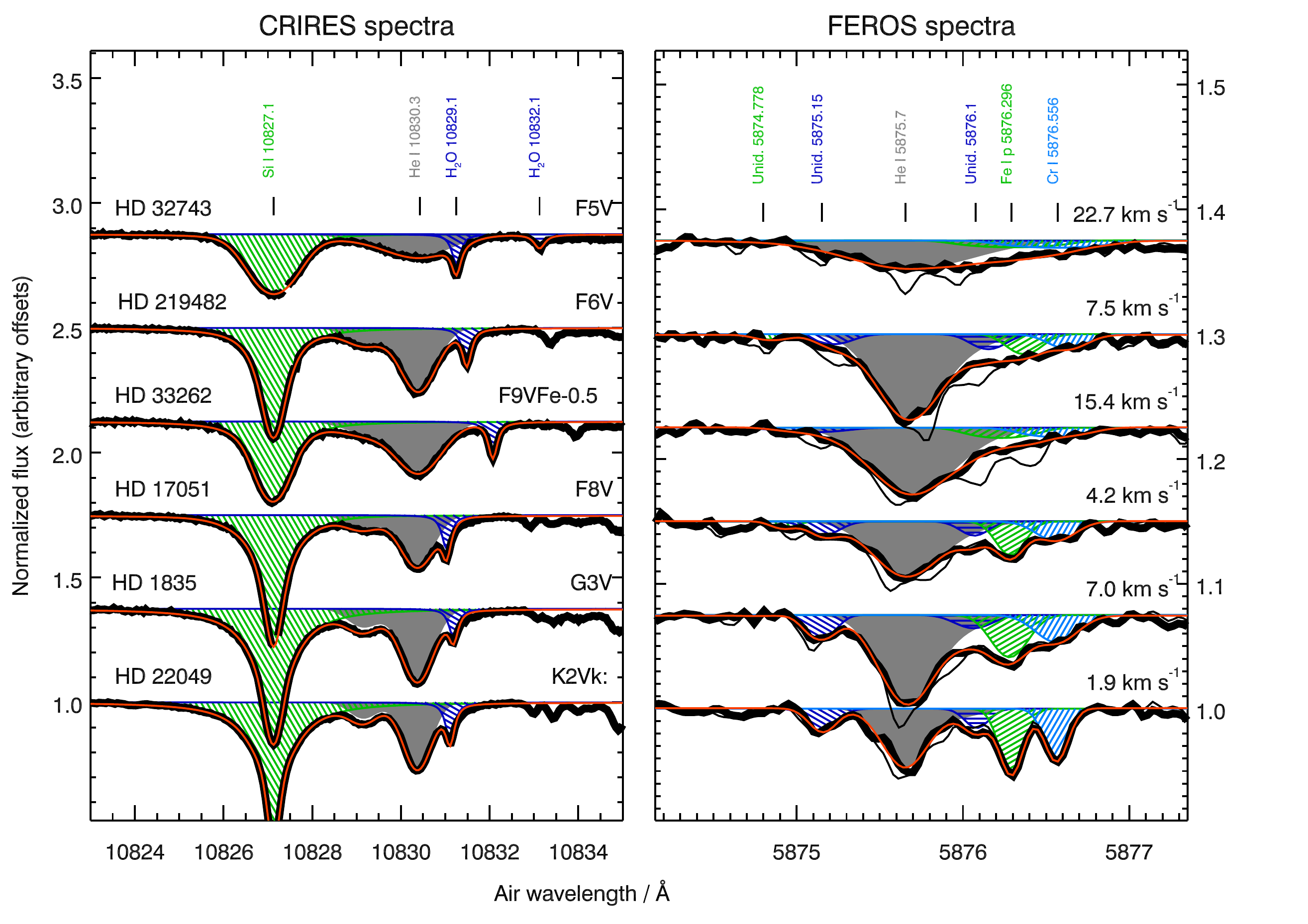}
    \caption{%
        Examples of observed \ion{He}{1} \heir\ (left) and \heiv\ (right)
        profiles and their fit functions. Normalized fluxes are offset by a constant value.  The
        fitted \ion{He}{1} multiplets are highlighted by a solid, grey area.
        The various nearby blends are highlighted by hatched areas.  The total
        fit function is represented in red.  In the panel showing the FEROS
        spectra, the thin, solid line shows the spectra before the telluric
        correction.  The spectral type and the \vsini\ value adopted to
        broaden the stellar lines are also indicated (in the right- and
        left-hand panels, respectively).  %
    }
    \label{fig:fits}
  \end{center}
\end{figure*}

  After the data were reduced following standard instrument
  pipelines, 
  we analyzed the 1D spectra around the wavelengths of
  interest (5876~\AA\ and 10830~\AA) using the Interactive Data Language
  (IDL).
  The spectra in these regions are shown in
  Fig.~\ref{fig:spectra}.
  In both wavelength ranges, it was necessary to correct the object spectra
  for contamination by terrestrial lines by using the spectra of A stars
  obtained in the same observing run.  Contamination by stellar blends also
  needed to be taken into account.  In some cases, further uncertainties in the procedure are
  introduced by rotational smearing.  Since the characteristics
  of telluric lines and stellar blends are different in the two wavelength
  ranges, the procedures for measuring the equivalent widths of the \ion{He}{1}
  lines are slightly different, as described in detail in the following two
  sections.  
  Examples of corrected spectra and of fitted line profiles according to the procedures described in
  the following sections are shown in Fig.~\ref{fig:fits}.
  A summary of the results together with relevant stellar parameters is given
  in Tables~\ref{tab:targets_a} and \ref{tab:targets_b}.
  In particular, we adopt the \vsini\ values given by \cite{vsini12}, when
  available, or by \cite{vsini05}, with the exception of HD~48189A, for which
  the value given by \cite{vsini09} was adopted. %

\subsection{Measuring the \dthree\ line}
\label{sec:reduction:spectra:D3}
  The \ion{He}{1} \heiv\ multiplet consists of 6 fine-structure lines arising
  from the transitions between levels $1s\,2p\: ^3$P and $1s\,3d\: ^3$D.  The
  rest wavelength of the strongest component is at 5875.615~\AA; four of the
  other components are separated at most by 25~m\AA\ from this component,
  while a sixth component, whose strength accounts for 1/9$^\mathrm{th}$ of
  the total (exact value in LS coupling), 
  is at 0.351~\AA\ on the red side, giving a slightly asymmetric
  shape to the line in high-quality spectra.

  The main difficulty in analyzing this line is that it is blended with a
  group of telluric H$_2$O lines at wavelengths of 5875.444~\AA,
  5875.596~\AA, 5875.769~\AA, 5876.124~\AA, and 5876.449~\AA, as listed by
  \cite{moore-etal:66}.  
    The procedure we adopted to correct for telluric
    blends relies on the observed spectra of telluric standards.
  Since the geometrical air mass inferred from the
  time of the observations is normally a poor indicator of the water vapor
  column mass for each spectrum, we chose instead to use as proxies the
  strongest H$_2$O lines in the range 5855--5930~\AA\ which appear to be not
  blended with stellar lines in all the spectra.  We found that the best proxy
  for this purpose is the sum of the equivalent widths of the H$_2$O lines at
  5919.6~\AA, 5920.6~\AA, and 5925.0~\AA.  From the measured value of this proxy in
  each object spectrum, we derived the corresponding telluric spectrum by
  interpolating the spectra of the telluric standard pixel-by-pixel as a
  function of the proxy.  This procedure of course ignores the details of the
  excitation of the individual H$_2$O lines and of all the other telluric lines.
    We estimated the error introduced by the telluric correction procedure on
    the \heiv\ line at its rest wavelength and for \vsini\ = 0 to be 1~m\AA\ rms
    or less; the error increases to 2.7 and 5.0~m\AA\ for profiles
    rotationally broadened by 40 and 80~km~s$^{-1}$ respectively. 
    Assuming the telluric correction to be proportional to the line
    broadening, we estimated its error correction as 
    $\sigma_\mathrm{atm} = 0.89 + 0.047\times\vsini$,
    with \vsini\ in km~s$^{-1}$ and $\sigma_\mathrm{atm}$ in m\AA. %
    After the telluric spectrum was removed, we fitted the \dthree\ line
    profile with a composite profile of 6 Gaussians, each representing one of
    the fine structure components of the multiplet.  
      The wavelength and Gaussian width of the reference component
      $\lambda5875.615$ were allowed to vary, while the wavelength separations
      and relative Gaussian widths of the other components were kept
      constant. 
    The relative strengths were also kept
    constant at values proportional to the relative $gf$ values.
      Thus, this composite profile for the \heiv\ multiplet is still
      determined by only three free parameters (multiplet equivalent width,
      position and width of the main component) as in standard,
      single-Gaussian fitting procedures. 
    The multiple-Gaussian profile thus constructed was then broadened by
    the \vsini\ value given in literature for each target (see
    Tables~\ref{tab:targets_a} and \ref{tab:targets_b}).

    The solar line list compilation by \cite{moore-etal:66} reports the
    presence of at least three photospheric lines potentially affecting the
    measurement of the \heiv\ equivalent width: \ion{Fe}{1} $\lambda$5876.30,
    \ion{Cr}{1} $\lambda$5876.45, and an unidentified line at 5874.778~\AA.
    In several spectra, particularly in cooler stars, we also noticed 
    at least two
    other unidentified lines at 5875.15~\AA\ and 5876.1~\AA\ 
    (see Fig.~\ref{fig:spectra}).  
    All these lines can affect the determination of the
    equivalent width of the \dthree\ line for $\vsini > 30$~km~s$^{-1}$.

    In nearly all the spectra, we included some or all of those lines in the
    fit procedure %
      as rotationally broadened Gaussians, adopting the
      \vsini\ value given in literature.  In spectra with significant
      rotational broadening we constrained the wavelengths and, in some cases,
      the Gaussian widths of the fitting functions within reasonable bounds to
      prevent unphysical results. %
      We found that these blends tend to increase towards cooler spectral
      types, thus confirming their stellar origin.  Only the very weak
      5874.778~\AA\ blend does not exhibit an obvious trend with \BV.  
        Table~\ref{tab:BV_fits} reports the coefficients of the linear fit of
        the equivalent widths of these stellar blends, measured in m\AA, as
        functions of \BV\ in the form:
        $A_\mathrm{\BV}+B_\mathrm{\BV}\times\BV$, together with the standard
        deviation from the fit, $\sigma_\mathrm{\BV}$.
      For those stars with high \vsini\ for which some or all of those blends
      could not be measured, the above relation can be used to estimate their
      contribution to the \ion{He}{1} \heiv\ equivalent width.  In particular, we found
      that the equivalent widths of the \ion{Fe}{1}~$\lambda$5876.30 and
      \ion{Cr}{1} $\lambda$5876.45 lines increase with \BV\ from $\sim
      2$~m\AA\ at \BV=0.4 to about $\sim 12$ and $\sim 8$~m\AA\ at \BV=1,
      respectively.
\begin{table}[h!]
\renewcommand{\thetable}{\arabic{table}}
\tablenum{2}
\centering
\caption{%
    Linear fit coefficients for stellar blends near \ion{He}{1} \heiv%
} \label{tab:BV_fits}
\begin{tabular}{lDDD}
\tablewidth{0pt}
\hline
\hline
Line & \multicolumn2c{$A_\mathrm{\BV}$} & \multicolumn2c{$B_\mathrm{\BV}$} & \multicolumn2c{$\sigma_\mathrm{\BV}$} \\
     & \multicolumn2c{m\AA}           & \multicolumn2c{m\AA}    & \multicolumn2c{m\AA} \\
\hline
\decimals
Unid. $\lambda5875.778$        &   0.7  &   .  & 0.4 \\
Unid. $\lambda5875.15$         &  -0.4  &  4.7 & 1.1 \\
Unid. $\lambda5876.1$          &   0.9  &  1.8 & 0.7 \\
\ion{Fe}{1} $\lambda5876.296$  &  -1.8  & 12.9 & 1.8 \\
\ion{Cr}{1} $\lambda5876.556$  &  -1.3  &  9.4 & 1.3 \\
\hline
\end{tabular}
\end{table}

        Assuming a Poissonian noise model corresponding to the estimated SNR
        on the continuum, the resulting $\chi^2$ of the fits are of the order
        of unity while the corresponding, formal uncertainties on the \heiv\ 
        multiplet equivalent width, $\sigma_\mathrm{fit}$, are less than
        1~m\AA.  
 
        To estimate the effect of possible residual blends or of deviation
        from Gaussian profiles, we removed the fitted blending lines from the
        spectra.  We then computed the equivalent width, $W_\mathrm{int}$, of
        the \heiv\ multiplet by trapezoidal integration of the residual
        intensity.  We estimated the uncertainty due to residual undetected
        blends or deviations from Gaussian profiles as $\sigma_\mathrm{nG}^2 =
        (W_\mathrm{int}-W_\mathrm{fit})^2$, where $W_\mathrm{fit}$ is the
        equivalent width from the multi-Gaussian fit procedure described
        above. These differences are typically of the order of 1~m\AA, and in
        any case below 3~m\AA\ even in the faster rotators. %
      
        We then estimated the overall uncertainty on the measured \heiv\
        equivalent width by summing quadratically the above estimates with the
        uncertainty due to the removal of the telluric spectrum:
        $\sigma_\mathrm{tot}^2 = \sigma_\mathrm{fit}^2 + \sigma_\mathrm{nG}^2
        + \sigma_\mathrm{atm}^2$. The \ion{He}{1} \heiv\ equivalent width and
        its associated uncertainty thus estimated are reported as $W$ in
        the 5$^\mathrm{th}$ column of Tables~\ref{tab:targets_a} and
        \ref{tab:targets_b}.

        Finally, in those spectra for which not all the stellar blends could
        be fitted, we took advantage of the estimated equivalent widths of
        those lines from the coefficients listed in Table~\ref{tab:BV_fits}.
        The sum of the blends falling within the \ion{He}{1} \heiv\ line
        profile but not explicitly fitted, along with their
        uncertainties (from a quadratic sum), is reported as $W_\mathrm{bl}$
        in Tables~\ref{tab:targets_a} and \ref{tab:targets_b}.  The equivalent
        width of the \ion{He}{1} \heiv\ multiplet obtained by subtracting
        this estimated contribution is reported as $W_\mathrm{corr}$. 

\subsection{Measuring the \heir\ line}
\label{sec:reduction:spectra:10830}
  The \ion{He}{1} \heir\ triplet arises from the transitions between levels
  $1s\,2s\: ^3$S and $1s\,2p\: ^3$P.  The two principal components are usually
  observed as a single line whose rest wavelength is at 10830.34~\AA, since
  they are separated by only 90~m\AA; the third component is at 10829.09~\AA\
  and is therefore often resolved in solar and stellar spectra.  
  In optically thin conditions, the $gf$ value 
  of this minor component is 
    in the ratio 1:8 (exact value in LS coupling) relative to the sum of the
    other two components.

  As in the case of the \dthree\ line, there are some nearby telluric H$_2$O lines
  which could interfere with the measurement of equivalent widths of
  the line components
  \citep{swensson-etal:70,breckinridge-hall:73}.  The telluric line closest to
  the rest wavelength of the main component is at 10830.0~\AA, but is so weak
  we were unable to detect it even in the spectra of the A-type, telluric
  standards.  The next closest, easily discernible H$_2$O line is at
  10832.1~\AA. 
  Another H$_2$O line at 10834.0~\AA, while fainter, also had to be
  taken into account in a few cases. 
    These telluric lines were both fitted with two Gaussian profiles with
    fixed wavelengths.

    The red wing of the nearby, strong \ion{Si}{1} $\lambda$10827.14~\AA\
    line often affects the profile of the \ion{He}{1} \heir\ triplet.  The
    problem of treating this blend has been extensively discussed in the
    context of spatially resolved solar spectra
    \citep{giovanelli-hall:77,jones:03,malanushenko-jones:04}.  In these high
    SNR spectra without evident velocity shifts, however, we found that the
    wings of the \ion{Si}{1} line could satisfactorily be fitted by a Voigt
    profile. We also found that fits could be improved if the core of the
    \ion{Si}{1} line is fitted separately with a Gaussian profile with the
    same central wavelength.

      Finally, the \ion{He}{1} \heir\ triplet was represented by three
      Gaussians at 10830.34~\AA, 10830.25~\AA, and 10829.09~\AA\ whose
      wavelength separations and relative Gaussian widths were kept constant.
      The relative strengths of the components at 10830.34~\AA\ and
      10830.25~\AA\ were also fixed at the 5:3 ratio of their $gf$ values.  

  The fitting functions for the stellar lines were broadened using the \vsini\
  value from literature (see Tables~\ref{tab:targets_a} and \ref{tab:targets_b}).
  In some spectra, the large value of \vsini\
  prevented the independent fit of the strength of the 
  \ion{He}{1} minor component with respect to the other two;
   in this case, the 
     relative strength of the \ion{He}{1} $\lambda$10829.09~\AA\ to the sum of
     the other two was fixed to the optically thin ratio of 1:8. %
     In summary, the \ion{He}{1} \heir\ triplet is fitted with a composite,
     multi-Gaussian profile with three or four free parameters: multiplet
     equivalent width, position and width of the main component, and, when
     possible, the equivalent width of the minor component at 10829.09~\AA.

     As in the case of the FEROS spectra, we assumed a Poissonian noise model
     corresponding to the estimated SNR on the continuum.  The resulting
     $\chi^2$ of the fits are of the order of unity while the corresponding,
     formal uncertainties on the \heir\ multiplet equivalent width,
     $\sigma_\mathrm{fit}$, are normally less than 3~m\AA, with some exceptions
     reaching 7 m\AA.

    In contrast with the determination of the \dthree\ equivalent width, the
    determination of the nearby stellar \ion{Si}{1} line and the removal of
    the telluric H$_2$O lines do not dominate the error budget.  

      We however estimated the effect of possible residual blends or 
      deviations from Gaussian profiles adopting the same approach as for the FEROS
      spectra.  We thus obtained an estimate of the uncertainty due to residual
      undetected blends or deviations from Gaussian profiles as
      $\sigma_\mathrm{nG}^2 = (W_\mathrm{int}-W_\mathrm{fit})^2$, from the
      spectra after the telluric and stellar blends were removed.  These
      differences are below 10~m\AA\ (in average of the order of 5~m\AA). The
      only exception is HD~17925, probably due to a CN blend at 10831.37~\AA\
      included in the integration of the residual profile but effectively
      filtered out by the fit procedure.  In the case of the \ion{He}{1}
      $\lambda$10829.09~\AA\ component, we scaled the corresponding estimate
      as
      $\sigma_\mathrm{fit}(\lambda10829.09)/\sigma_\mathrm{fit}(\mathrm{total})$.
      The overall uncertainties from the fit procedure listed in
      Tables~\ref{tab:targets_a} and ~\ref{tab:targets_b} are then obtained as
      $\sigma_\mathrm{tot}^2 = \sigma_\mathrm{fit}^2 + \sigma_\mathrm{nG}^2$.

\section{Results and discussion}
\label{sec:results}

The list of objects with our equivalent width measurements of the triplet
lines is given in Tables~\ref{tab:targets_a} and \ref{tab:targets_b}.
  In the case of the \dthree\ line, we report both the equivalent width
  measured as described in the previous section, $W_\mathrm{fit}$(\ion{He}{1}
  \heiv), and the corrected value, $W_\mathrm{corr}$(\ion{He}{1} \heiv),
  obtained after an estimate of the sum of residual, undetected stellar blends
  in the line profile is removed, also reported in those tables as
  $W_\mathrm{bl}$(\heiv).
Additional properties of the stars
that are relevant to this investigation are included in
Tables~\ref{tab:targets_a} and \ref{tab:targets_b}, as well as the determination of filling
  factors, $f$, derived from Eq.~\ref{eq:Wobs} 
  for the two series of models
  we have considered, together with their estimated errors determined as
  described in Sec.~\ref{sec:theory} and illustrated in
  Fig.~\ref{fig:joint_response}.
  Tables~\ref{tab:targets_a} and \ref{tab:targets_b} also list the minimum filling factors,
  $f_\mathrm{min}$, from Eq.~\ref{eq:fmin}, %
  obtained from the measurement of the \heir\ line only, using the values
  $\Wq=40$~m\AA\ and $\Wmax=410$~m\AA\ discussed in
  Sec.~\ref{sec:theory:atmo}.

\floattable
\begin{deluxetable*}{lccccccccccc}
\tablenum{3}
\tabletypesize{\scriptsize}
\rotate
\tablewidth{0pt}
\tablecaption{Helium Equivalent Width Measurements and Data for Program Stars: $\bv\le 0.5$}
\tablehead{
\colhead{HD} & \colhead{\BV} & \colhead{\vsini\tablenotemark{a}} & \colhead{$\log L_X$\tablenotemark{b}} & \colhead{$W_\mathrm{fit}$(\ion{He}{1} $\lambda$5876)} & \colhead{$W_\mathrm{corr}$(\ion{He}{1} $\lambda$5876)} & \colhead{$W_\mathrm{bl}$($\lambda$5876)} & \colhead{$W$(\ion{He}{1} $\lambda$10830)} & \colhead{$W$(\ion{He}{1} $\lambda$10829.1)} & \colhead{$f$(C)} & \colhead{$f$(C-np)} & \colhead{$f_\mathrm{min}$} \\
\colhead{} & \colhead{} & \colhead{km~s$^{-1}$} & \colhead{erg s$^{-1}$} & \colhead{m\AA} & \colhead{m\AA} & \colhead{m\AA} & \colhead{m\AA} & \colhead{m\AA} & \colhead{} & \colhead{} & \colhead{}
}
\startdata
HD 49933   &       0.36 &        9.9 &      29.50 &       27.6(1.4) &       27.6(1.4) &         \nodata &         \nodata &         \nodata &         \nodata &         \nodata &    \nodata \\
HD 29992   &       0.37 &       97.5 &      28.83 &       27.1(5.5) &       18.4(6.1) &        8.7(2.6) &      364.3(9.6) &         \nodata &         \nodata &         \nodata &       0.88 \\
HD 219693  &       0.39 &       19.9 &      28.98 &       30.1(1.8) &       26.4(2.3) &        3.7(1.4) &      259.6(5.8) &       38.5(5.1) &      0.66(0.04) &      0.72(0.03) &       0.59 \\
HD 32743   &       0.39 &       22.7 &      29.02 &       21.3(2.2) &       18.3(2.5) &        3.0(1.3) &      184.2(4.2) &       28.6(3.7) &      0.44(0.04) &      0.50(0.03) &       0.39 \\
HD 104731  &       0.41 &       15.9 &      28.49 &       14.1(1.7) &       10.3(2.2) &        3.8(1.4) &         \nodata &         \nodata &         \nodata &         \nodata &    \nodata \\
HD 3302    &       0.41 &       17.8 &      29.40 &       33.2(2.0) &       30.1(2.4) &        3.1(1.3) &      305.0(6.0) &       48.7(5.3) &      0.83(0.05) &      0.89(0.04) &       0.72 \\
HD 77370   &       0.42 &       60.4 &      29.07 &       23.2(4.6) &       16.6(4.9) &        6.6(1.9) &         \nodata &         \nodata &         \nodata &         \nodata &    \nodata \\
HD 68456   &       0.43 &        8.8 &      29.15 &       33.1(1.3) &       31.4(1.5) &        1.7(0.7) &         \nodata &         \nodata &         \nodata &         \nodata &    \nodata \\
HD 30652   &       0.44 &       17.3 &      29.03 &       20.8(2.0) &       17.4(2.4) &        3.3(1.3) &     187.7(11.0) &       31.6(9.1) &      0.47(0.07) &      0.52(0.06) &       0.40 \\
HD 88697   &       0.44 &       19.8 &      29.49 &       38.1(2.1) &       34.1(2.5) &        4.0(1.4) &         \nodata &         \nodata &         \nodata &         \nodata &    \nodata \\
HD 76653   &       0.45 &       10.3 &      29.33 &       28.8(1.4) &       27.1(1.6) &        1.7(0.7) &         \nodata &         \nodata &         \nodata &         \nodata &    \nodata \\
HD 201647  &       0.45 &       22.4 &      28.92 &       18.5(2.1) &       14.5(2.5) &        4.1(1.4) &      174.7(9.1) &       24.6(8.1) &      0.45(0.07) &      0.50(0.07) &       0.36 \\
HD 79940   &       0.45 &      117.2 &      28.79 &       28.5(6.9) &       17.4(7.4) &       11.0(2.6) &         \nodata &         \nodata &         \nodata &         \nodata &    \nodata \\
HD 37495   &       0.46 &       27.2 &      29.31 &       25.9(2.2) &       21.7(2.6) &        4.1(1.4) &      290.2(8.6) &         \nodata &      0.89(0.07) &      0.92(0.05) &       0.68 \\
HD 219482  &       0.47 &        7.5 &      29.42 &       35.9(1.3) &       35.9(1.3) &         \nodata &      318.5(4.3) &       62.6(3.6) &      0.85(0.03) &      0.91(0.02) &       0.75 \\
HD 33262   &       0.47 &       15.4 &      28.71 &       34.9(1.6) &       33.1(1.8) &        1.7(0.7) &      317.2(4.3) &       56.3(3.7) &      0.86(0.04) &      0.92(0.03) &       0.75 \\
HD 189245  &       0.49 &       72.6 &      29.90 &       76.9(5.3) &       64.7(5.9) &       12.2(2.6) &      409.6(4.1) &         \nodata &         \nodata &         \nodata &       1.00 \\
HD 199260  &       0.50 &       13.7 &      29.18 &       31.4(1.8) &       29.5(2.1) &        1.9(1.1) &      294.3(5.7) &       51.5(4.9) &      0.81(0.05) &      0.87(0.04) &       0.69 \\
\enddata
\tablenotetext{a}{From \cite{vsini12} when available, otherwise from \cite{vsini05}.}
\tablenotetext{b}{From \cite{ROSATnearby99}, except $L_\mathrm{X}$ for HD~49933 which is from \cite{ROSATbright98}.}
\label{tab:targets_a}
\end{deluxetable*}

\floattable
\begin{deluxetable*}{lccccccccccc}
\tablenum{4}
\tabletypesize{\scriptsize}
\rotate
\tablewidth{0pt}
\tablecaption{Helium Equivalent Width Measurements and Data for Program Stars: $\bv> 0.5$}
\tablehead{
\colhead{HD} & \colhead{\BV} & \colhead{\vsini\tablenotemark{a}} & \colhead{$\log L_X$\tablenotemark{b}} & \colhead{$W_\mathrm{fit}$(\ion{He}{1} $\lambda$5876)} & \colhead{$W_\mathrm{corr}$(\ion{He}{1} $\lambda$5876)} & \colhead{$W_\mathrm{bl}$($\lambda$5876)} & \colhead{$W$(\ion{He}{1} $\lambda$10830)} & \colhead{$W$(\ion{He}{1} $\lambda$10829.1)} & \colhead{$f$(C)} & \colhead{$f$(C-np)} & \colhead{$f_\mathrm{min}$} \\
\colhead{} & \colhead{} & \colhead{km~s$^{-1}$} & \colhead{erg s$^{-1}$} & \colhead{m\AA} & \colhead{m\AA} & \colhead{m\AA} & \colhead{m\AA} & \colhead{m\AA} & \colhead{} & \colhead{} & \colhead{}
}
\startdata
HD 41700   &       0.52 &       14.7 &      29.63 &       45.6(1.6) &       41.1(2.1) &        4.5(1.4) &         \nodata &         \nodata &         \nodata &         \nodata &    \nodata \\
HD 100563  &       0.53 &       13.5 &      29.13 &       26.2(1.6) &       21.6(2.1) &        4.6(1.4) &         \nodata &         \nodata &         \nodata &         \nodata &    \nodata \\
HD 17051   &       0.57 &        4.2 &      28.83 &       24.2(1.1) &       24.2(1.1) &         \nodata &      239.0(1.4) &       47.7(1.1) &      0.66(0.03) &      0.71(0.03) &       0.54 \\
HD 88742   &       0.59 &        2.7 &      28.58 &       15.8(1.0) &       15.8(1.0) &         \nodata &         \nodata &         \nodata &         \nodata &         \nodata &    \nodata \\
HD 48189A  &       0.61 &       15.5 &      29.99 &       38.3(1.6) &       35.8(2.0) &        2.5(1.1) &         \nodata &         \nodata &         \nodata &         \nodata &    \nodata \\
HD 30495   &       0.64 &        2.9 &      28.83 &       19.8(1.0) &       19.8(1.0) &         \nodata &      223.6(5.7) &       41.4(4.6) &      0.72(0.08) &      0.76(0.07) &       0.50 \\
HD 16417   &       0.66 &        2.5 &      27.81 &        4.5(1.2) &        4.5(1.2) &         \nodata &       71.7(9.0) &       12.1(7.3) &      0.23(0.18) &      0.31(0.20) &       0.09 \\
HD 1835    &       0.66 &        7.0 &      28.99 &       33.2(1.2) &       33.2(1.2) &         \nodata &      323.7(5.8) &       67.0(4.8) &      0.92(0.04) &      0.96(0.03) &       0.77 \\
HD 20630   &       0.67 &        4.5 &      28.89 &       27.4(1.1) &       27.4(1.1) &         \nodata &      279.0(4.8) &       55.6(3.7) &      0.82(0.05) &      0.87(0.04) &       0.65 \\
HD 76151   &       0.67 &        2.4 &      28.33 &       12.6(1.0) &       12.6(1.0) &         \nodata &         \nodata &         \nodata &         \nodata &         \nodata &    \nodata \\
HD 42807   &       0.68 &        3.6 &      28.68 &       24.1(1.1) &       24.1(1.1) &         \nodata &         \nodata &         \nodata &         \nodata &         \nodata &    \nodata \\
HD 43162   &       0.68 &        5.5 &      29.13 &       31.5(1.2) &       31.5(1.2) &         \nodata &         \nodata &         \nodata &         \nodata &         \nodata &    \nodata \\
HD 10700   &       0.72 &        1.8 &      26.30 &        3.2(1.2) &        3.2(1.2) &         \nodata &       51.6(3.0) &         \nodata &      0.12(0.14) &      0.19(0.15) &       0.03 \\
HD 17925   &       0.86 &        4.8 &      29.08 &       28.1(1.1) &       28.1(1.1) &         \nodata &     317.6(20.9) &      70.0(16.0) &         \nodata &         \nodata &       0.75 \\
HD 22049   &       0.88 &        1.9 &      28.32 &       18.1(1.0) &       18.1(1.0) &         \nodata &      257.8(3.6) &       51.3(3.0) &         \nodata &         \nodata &       0.59 \\
HD 5133    &       0.94 &        1.8 &      27.79 &       12.6(1.0) &       12.6(1.0) &         \nodata &     229.8(11.7) &       42.3(9.6) &         \nodata &         \nodata &       0.51 \\
\enddata
\tablenotetext{a}{From \cite{vsini12} when available, otherwise from \cite{vsini05} or (HD~48189A) from \cite{vsini09}.}
\tablenotetext{b}{From \cite{ROSATnearby99}.}
\label{tab:targets_b}
\end{deluxetable*}

\begin{figure*}
  \includegraphics[angle=90,width=1.05\linewidth]{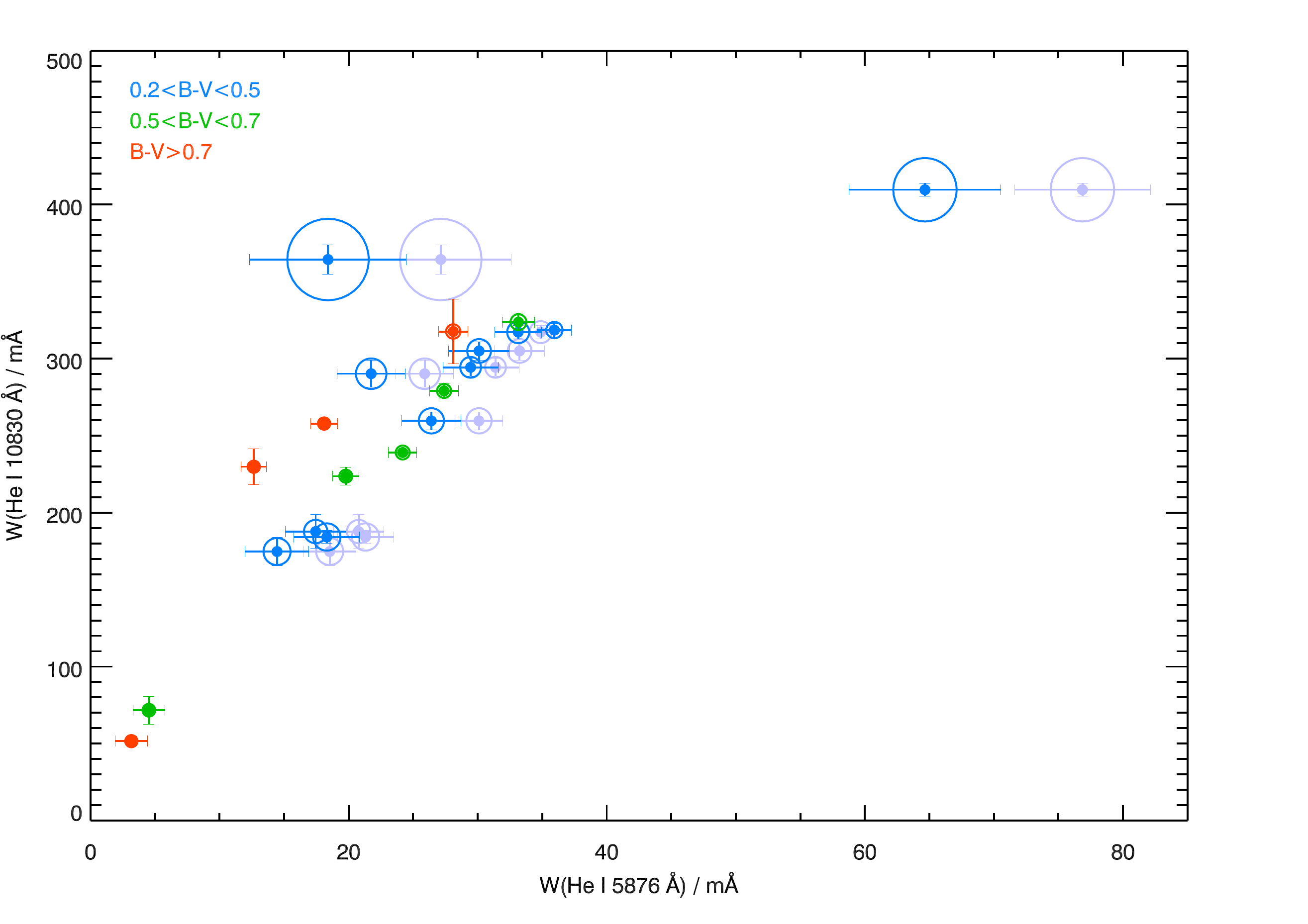}
  \caption{Observations of the \ion{He}{1} triplet lines in solar-type
    stars. 
      Error bars represent 1-$\sigma$ uncertainties, estimated as described in
      Sec.~\ref{sec:reduction:spectra}. 
      Points with an outer circle represent stars with a projected
      rotational velocity exceeding 15 km s$^{-1}$; the radius of the circle
      is proportional to \vsini. %
      Ranges in \BV\ are coded in colors. 
        Points corresponding to the \dthree\
        measurements not corrected for residual stellar blends,
        $W_\mathrm{fit}$(\ion{He}{1} \heiv), are shown in lighter color.
      The general correlation between \heir\ and \dthree\ is evident
      in these data. %
  }
  \label{fig:joint_eqws}
\end{figure*}

\begin{figure}
  \begin{center}
    \includegraphics[angle=90,width=1.05\linewidth]{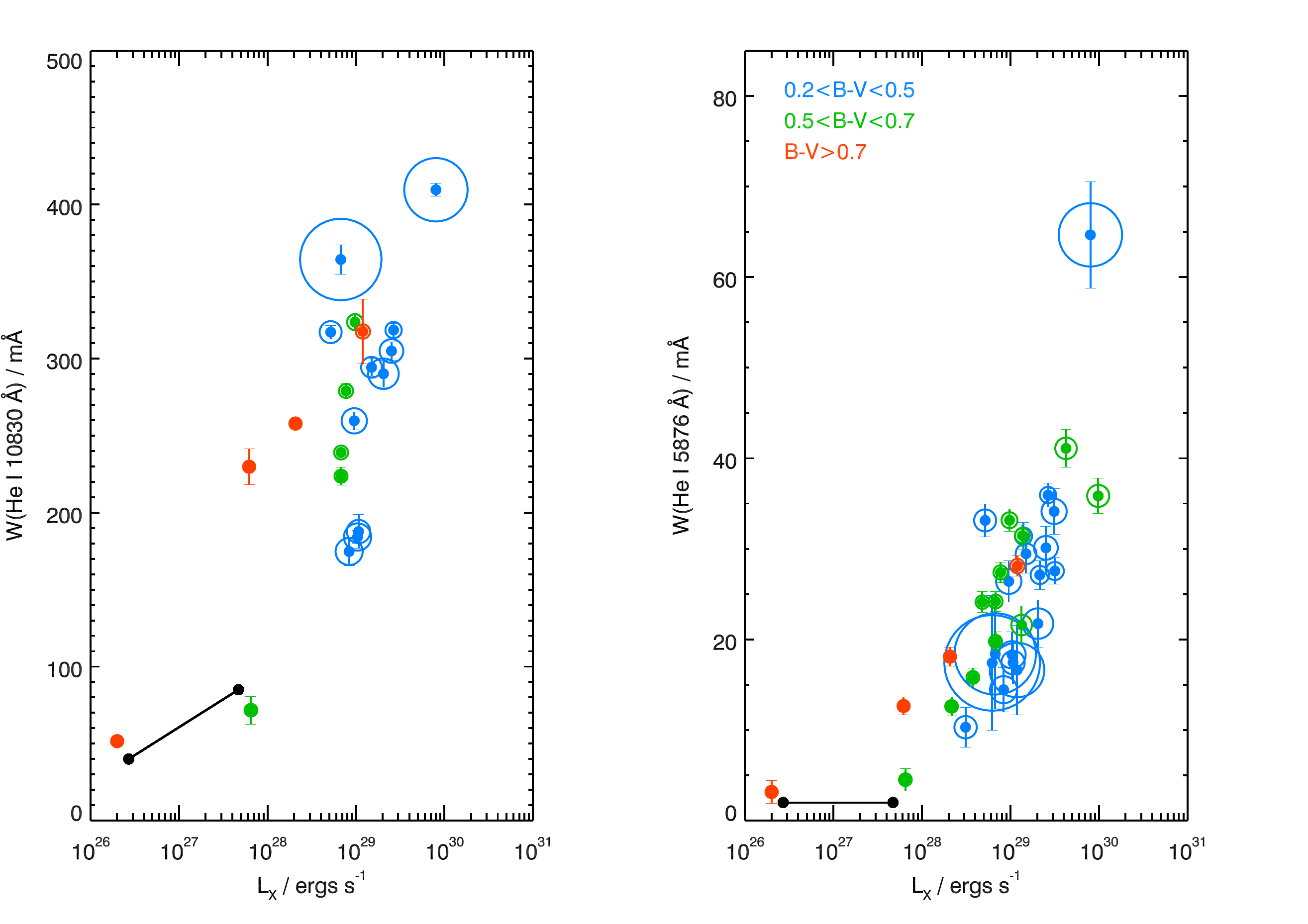}
    \caption{Strength of the individual triplet features vs. X-ray luminosity in
      our sample of solar-type stars.  
      Left: %
      The correlation between the equivalent
      width of \heir\ and coronal X-ray emission in this sample. %
      Right: %
      Similarly for the \dthree\ line at 5876~\AA. 
        The range of variability of the X-ray luminosity of the
        Sun 
        \citep{peres-etal:00} 
        and the corresponding Sun-as-a-star \heir\ equivalent widths
        \citep{harvey-livingston:94,livingston-etal:10} 
        are also shown as a black segment.
        We adopted an upper limit for the \heiv\ equivalent width of 2~m\AA.
      A possible saturation may be
      present in the \heir\ -- X-ray relationship. 
        This is expected from the discussion of the origin of
          $\Wmax$ for that line (Sec.~\ref{sec:theory}). 
    }
    \label{fig:eqws_vs_X}
  \end{center}
\end{figure}

  We display in Fig.~\ref{fig:joint_eqws} the correlation between \ion{He}{1}
  \heir\ and \dthree\ based on the data given in Tables~\ref{tab:targets_a}
  and \ref{tab:targets_b}.  %
    The data points corresponding to values $W_\mathrm{fit}$(\ion{He}{1}
    \heiv), when different from $W_\mathrm{corr}$(\ion{He}{1} \heiv), are
    shown in lighter colors. In the remainder of the discussion and in all
    the other figures, we will refer to $W_\mathrm{corr}$(\ion{He}{1} \heiv)
    simply as $W$(\ion{He}{1} \heiv).
In both cases, the expected
correlation between these diagnostics is evident in Fig.~\ref{fig:joint_eqws} though there
appears to be a saturation in the \heir\ feature at $W_{\lambda10830}\approx$
300 m{\AA}.  This saturation also is apparent in the correlation of
\heir\ with
X-ray luminosity as seen in the left-hand panel of
  Fig.~\ref{fig:eqws_vs_X}.
This is precisely the behavior expected on the basis of the
theoretical considerations discussed in Sec.~\ref{sec:theory}.
Likewise, the strength of the
{\dthree} feature is directly correlated with X-ray luminosity though no
saturation in this less optically thick line appears
(right-hand panel of Fig.~\ref{fig:eqws_vs_X}).  
We note,
parenthetically, that the correlations in Fig.~\ref{fig:eqws_vs_X} with
coronal X-ray emission
levels are not only the direct result of a photoionization-recombination line
formation process, as has been discussed by AG95 and AJ97 \citep[also
see][]{agj95}, but also incidental to the generally elevated degree of
magnetic activity and the corresponding enhanced average chromospheric and
coronal pressures.

\begin{figure}
  \begin{center}
    \includegraphics[angle=90,width=1.05\linewidth]{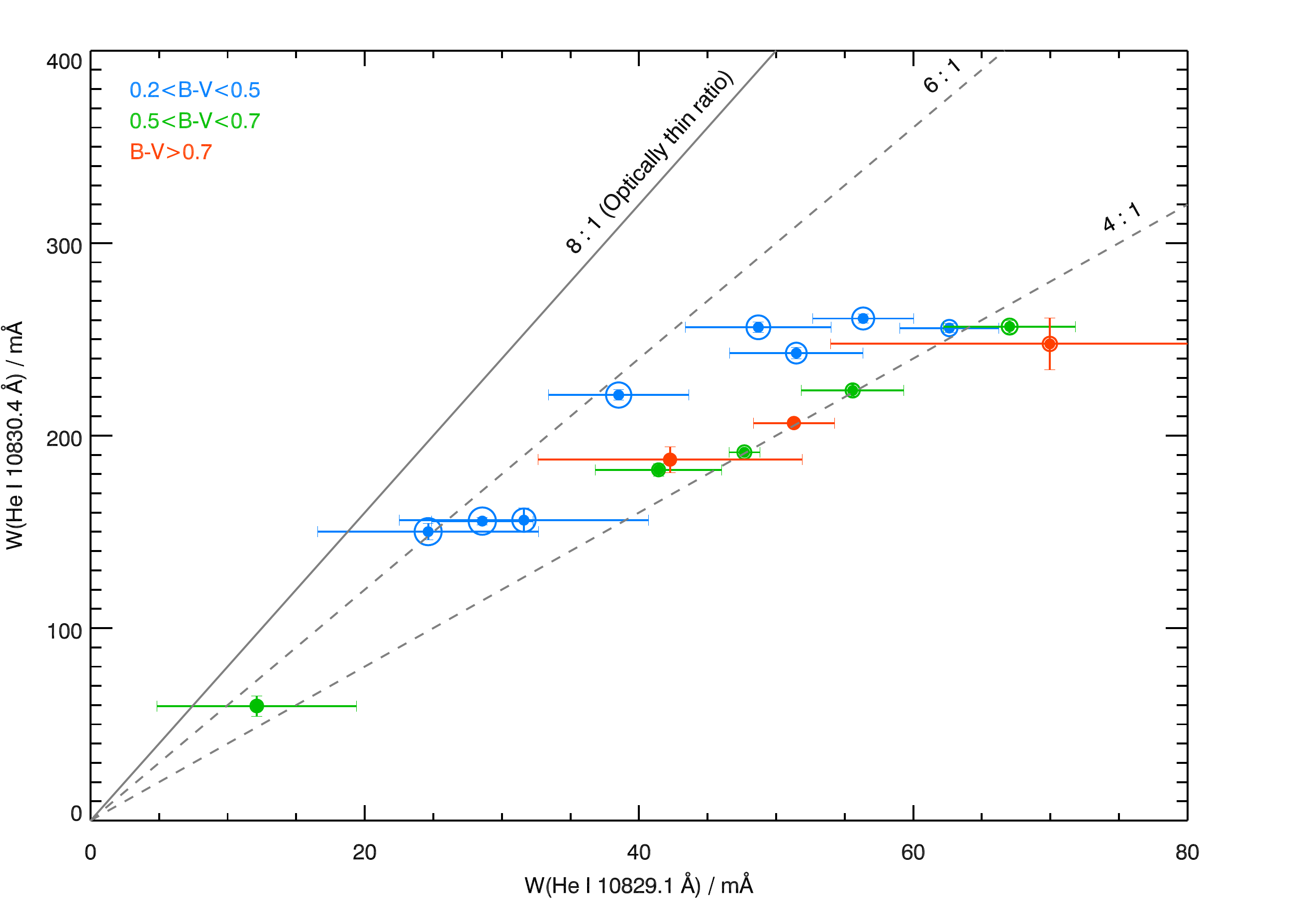}
    \caption{%
        Correlation of the main components of the \ion{He}{1} triplet at
        10830.4~\AA\ and the minor component at 10829.1~\AA.  The optically
        thin ratio (8:1) and two other representative values for the ratio of
        these fine structure components (6:1 and 4:1) are
        also shown.
    }
    \label{fig:eqws_triplet}
  \end{center}
\end{figure}

Figure~\ref{fig:eqws_triplet} shows the correlation between the minor
component of the \ion{He}{1} \heir\ triplet and the sum of the other two main
components at 10830.4~\AA\ for those stars for which the minor component at
10829.1~\AA\ could be fitted independently.  In those cases, the ratio between
the fine structure components range between 4 and 6, below the optically
thin limit of 8, suggesting a significant optical thickness of the main
component.  These values of the line ratio
are consistent with the stronger line profiles shown in
Fig.~\ref{fig:solar_spectra}.

\begin{figure*}
  \epsscale{1.0}
  \includegraphics[angle=90,width=1.05\linewidth]{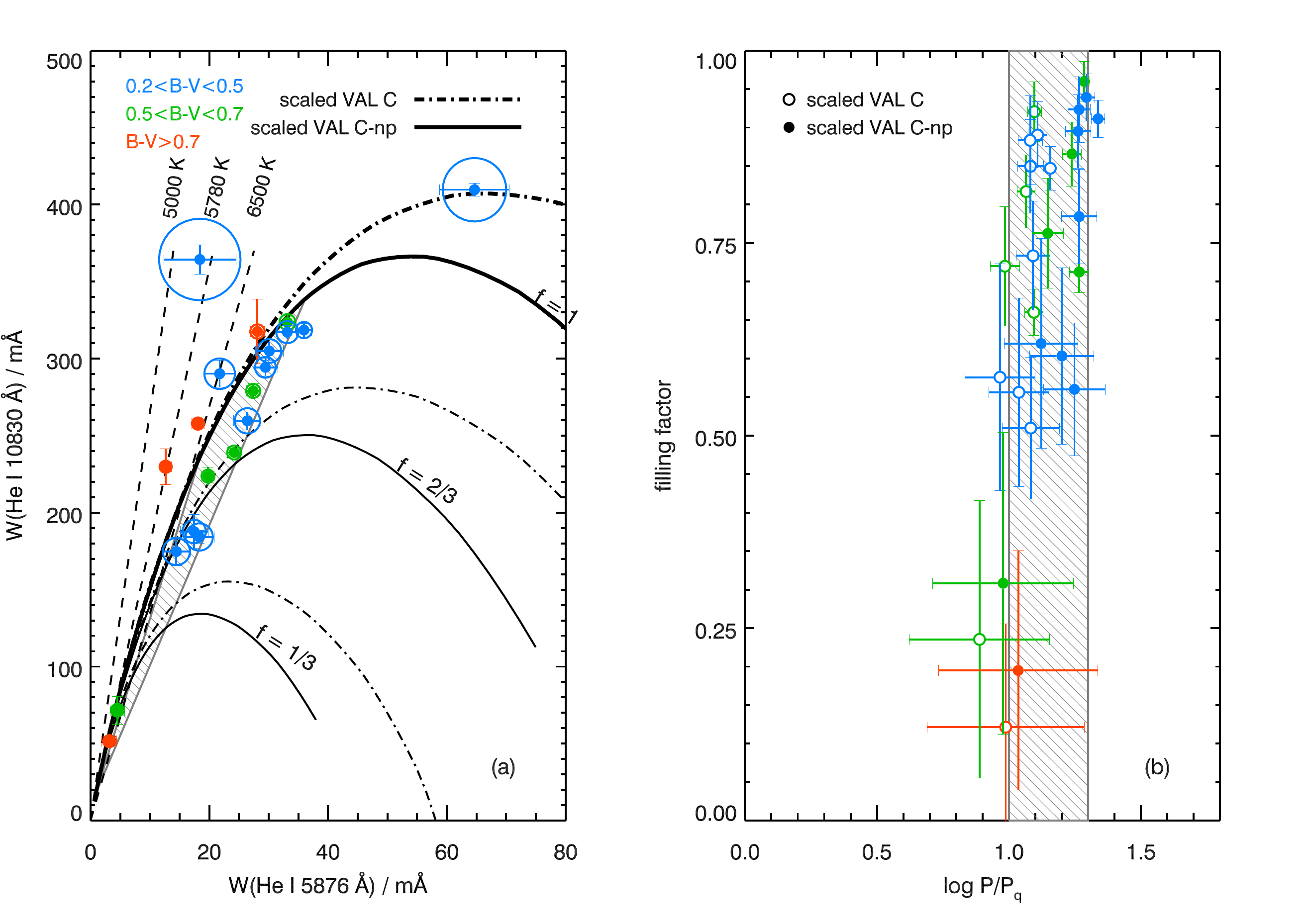}
  \caption{%
    Filling factors and chromospheric mass loading atop the chromosphere
    from observed helium line strengths. %
    Left-hand panel (a):  %
    Loci of fractional active region area coverage, or filling factor,
    are superimposed on the observed strengths of the helium triplet lines as
    measured in this stellar sample.  
    Data are represented with the same conventions as in
    Fig.~\ref{fig:joint_eqws} while theoretical models are represented
    with the same conventions as in Figures \ref{fig:joint_response} and
    \ref{fig:atmo_models}. 
      As in Fig.~\ref{fig:atmo_models}, the optically thin limit of the joint
      response of the two lines  as a function of stellar effective temperature is
      also shown. %
    Right-hand panel (b): 
    Filling factors and mass loading atop the chromosphere ($P/P_\mathrm{q}$)
    derived from the observed line strengths with the uncertainties estimated
    as described in Sec.~\ref{sec:theory}. The open and filled circles
    represent results from the C and C-np model series. %
    The dashed area in panel (b), corresponding to $1.0 < \log P/P_\mathrm{q}
    < 1.3$, is
    also mapped in panel (a) in the case of the C-np series. %
    See text for a discussion.
    }
    \label{fig:eqws_loci}
\end{figure*}

The measurements presented in Fig.~\ref{fig:joint_eqws} may be compared to the calculated joint
response of the two \ion{He}{1} triplet lines as given in the
left-hand panel of Fig.~\ref{fig:eqws_loci}. %
The results of the model calculations are given as
loci of filling factors for two different series of scaled chromospheric
models, 
C and C-np, 
described in Sec.~\ref{sec:theory:atmo} and shown in
Fig.~\ref{fig:atmo_models}.
  With a few exceptions discussed below, the majority of these simultaneous
  observations of {\dthree} and {\heir} appear below the model loci for unit
  filling factor in panel (a) of Fig.~\ref{fig:eqws_loci}.  
  In particular, the observed saturation in \heir\ follows the predicted
  behavior of the two lines.
  Note also that the error bars correspond to $1\sigma$ statistical uncertainties,
  and thus some of the points above the $f>1$ loci are consistent with the
  theoretical estimates within $2\sigma$ or $3\sigma$.
  We consider this
  as a confirmation that the approach is plausible, as already
  noted by AG95 by using less accurate, non-simultaneous data from the literature.

In particular, the
relatively high fractional area coverages generally exceeding $f \sim 0.3$ are
not surprising for active solar-type stars.  On the other hand, exceptions
occur in Fig.~\ref{fig:eqws_loci} with apparently unphysical filling factors $>$ 1.  The objects
with $\BV > 0.8$ are K dwarfs for which our dwarf G chromospheric models may
not apply.  Specifically, the background radiation temperatures associated
with the triplet transitions may be too high for K dwarf chromospheres, as
discussed in Sec.~\ref{sec:theory:atmo}.  
The dependence on \Teff\ of the joint response of the two lines is apparent
in the optically thin limit shown both in Fig.~\ref{fig:atmo_models} and
Fig.~\ref{fig:eqws_loci}.

An analogous argument does not apply to the early F dwarf exceptions that are
above the $f = 1$ locus in Fig.~\ref{fig:eqws_loci}
(HD~29992, HD~189245, and HD~37495),
since the higher
background photospheric radiation at 10830~\AA\ would have the effect of increasing the gap between
the data points and the $f = 1$ locus, 
though additional effects merit consideration
in these cases.  
In particular, 
  the procedure to account for undetected stellar blends described in
  Sec.~\ref{sec:reduction:spectra:D3} may be over-correcting the strength of the
  \dthree\ in these relatively rapid rotators (see Fig.~\ref{fig:joint_eqws}).
In addition, non-magnetic contributions to the triplet line strengths may be
present that include (a) a non-negligible photospheric contribution in these
relatively warmer stars and (b) a contribution by acoustic heating due to the
presence of more vigorous convective velocities.  The former can be corrected
through the application of appropriate photospheric models while the latter
may yield insight on the role of acoustic heating in stellar chromospheres.

  For each pair of equivalent width measurements falling within the allowed
  region of the $(\dthree,\heir)$ plane, the corresponding pair of
  parameters $(f,P/P_\mathrm{q})$ derived as described in
  Sec.~\ref{sec:theory} is shown in the right-hand panel of
  Fig.~\ref{fig:eqws_loci}. The results derived from the two adopted series
  of models, C and C-np, are shown with open and filled circles, respectively.

  We recall that the adoption of the appropriate background photospheric model
  in the computation of the \ion{He}{1}
  triplet lines would likely lower the equivalent width of the \heir\ line with
  respect to that for the \dthree\ feature. We would thus expect the theoretical joint response
  corresponding to $f=1$ to fall closer to the observed data points for
  $0.2<\BV<0.5$. For these objects, therefore, the filling factors shown in
  the right-hand panel of Fig.~\ref{fig:eqws_loci} are probably biased
  towards lower values. %

\begin{figure}
\includegraphics[angle=90,width=1.05\linewidth]{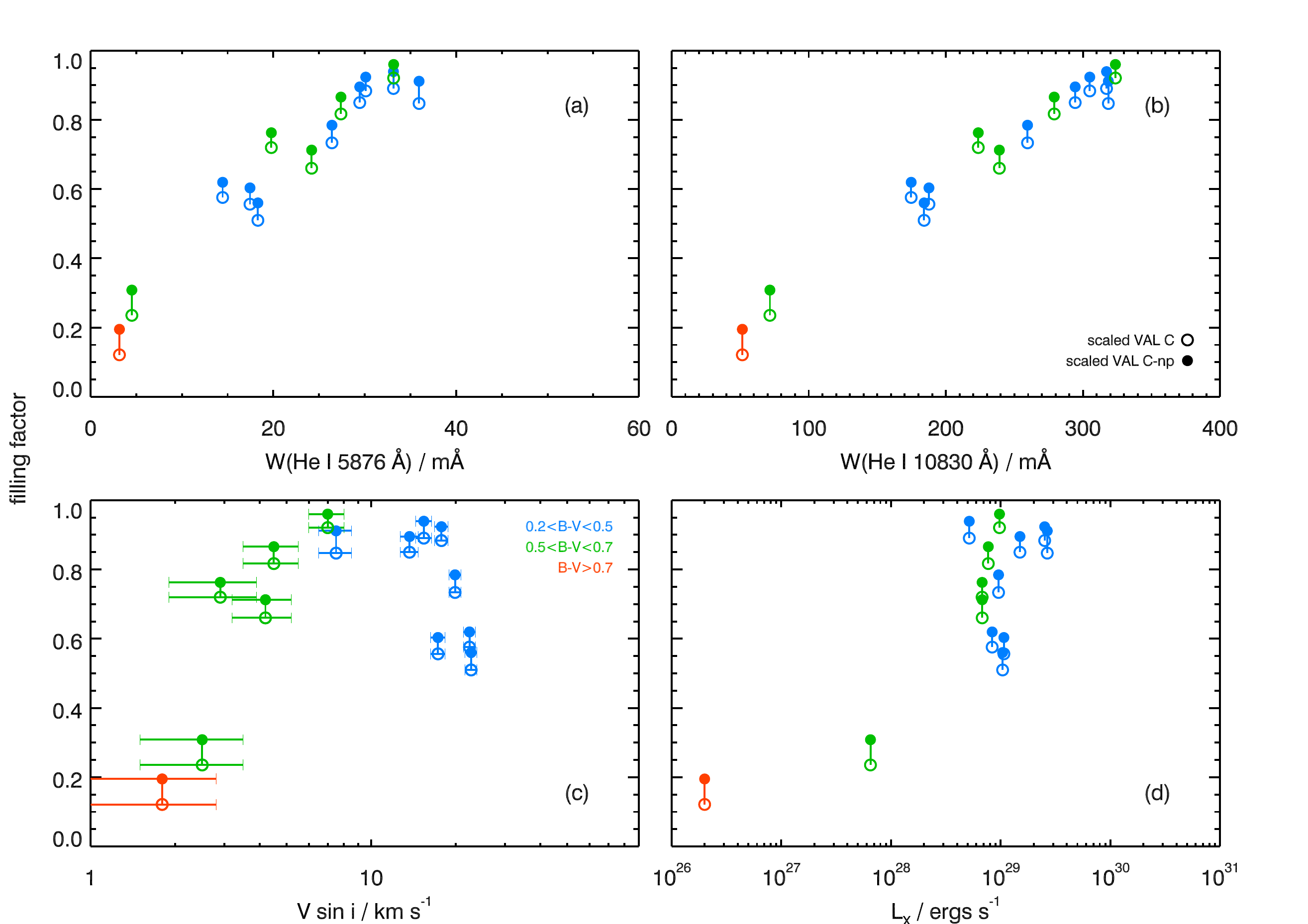}
\caption{%
  The calculated area coverage of active regions versus observed line
  strengths and stellar parameters, respectively. %
  The upper panels (a and b) show the
  inferred filling factor versus the equivalent widths measured for each
  helium triplet line. %
  Bottom panels: The deduced fractional area coverage versus
  projected stellar rotational velocity (c) and X-ray luminosity (d).
  The open and filled circles represent the two results from application of
  the two chromospheric models utilized in this investigation, as indicated in
  the upper right panel and following the convention used in
    Fig.~\ref{fig:eqws_loci}.  
  See text for a discussion.
  }
  \label{fig:eqws_ff}
\end{figure}

In Fig.~\ref{fig:eqws_ff} we apply the inferred filling factors to examine correlations
relevant to dynamo modeling and chromospheric heating processes.  
In that figure, we note that the main uncertainty in determining the
  filling factor comes from the choice of the specific set of models, even 
  neglecting the effect of the photospheric background described above. 
  Thus, in the remainder of this paper we 
  do not further discuss the errors shown in Fig.~\ref{fig:eqws_loci} 
propagated from the observational uncertainties.  
  On the other hand, since errors on rotational velocities can sometimes be
  substantial, we do show error bars on \vsini\ in panel (c).  We have assumed
  that typical uncertainties are not smaller than 5\% of \vsini, with a lower
  limit of 1 km~s$^{-1}$ for $\vsini<20$~km~s$^{-1}$.  We adopted these lower
  limits if no larger uncertainty is specified in the source reference.  We
  believe these lower limits are reasonable estimates because most of the
  uncertainties reported in \vsini\ measurements are of formal nature and do
  not normally reflect our knowledge on all broadening mechanisms involved.

  Inspection of Fig.~\ref{fig:eqws_ff} seems to indicate an interesting
  correlation between X-ray luminosity and the surface area coverage (panel
  d). 
  We recall that \cite{pizzolato-etal:03} found the onset of saturation near
  $L_\mathrm{X} \sim 10^{30}$ ergs s$^{-1}$ for solar-type stars in the broad
  range of rotation periods of $\sim$ a few to approximately 10 days. 
  Our Fig.~\ref{fig:eqws_ff}(d) would suggest that this saturation may
  correspond to active region filling factors of $\sim 0.8$ -- $1.0$.

  A relation between projected rotational velocity and active region filling
  factor in this small sample of stars is less evident (panel c of
  Fig.~\ref{fig:eqws_ff}).  Nevertheless, in the case of the cooler stars
  there is a trend of increasing filling factor with \vsini.  In the case of
  hotter stars, the trend seems to be opposite, perhaps indicating a
  saturation of filling factors in the range 10--20~km~s$^{-1}$.  Once again,
  we cannot exclude a bias in the determination of $f$ for these hotter
  objects.  
The empirical correlation in the upper panels of Fig.~\ref{fig:eqws_ff} merits further
investigation because of its potentially important implications for the nature
of chromospheric heating in active, solar-type stars.  We have from Eq.~\ref{eq:Wobs}
for $\Wq\approx 0$ that $f = \Wobs/\Wa(p)$, which yields a linear
relation for a constant slope of $1/\Wa(p)$.  Consequently, the results in
Fig.~\ref{fig:eqws_ff}a and b suggest that the intrinsic chromospheric heating in
surface active regions in solar-type stars is similar.  
We indeed find that the activity parameters $p=P/P_\mathrm{q}$ obtained by inverting
Eq.~\ref{eq:Wobs} for both \heiv\ and \heir\ typically fall in the range of
$10-30$, i.e., the active region densities are about an order of magnitude
higher than in the quiet Sun in the region of formation of the \ion{He}{1}
lines. A density contrast of this magnitude is consistent, for instance, with
the model calculations by
\cite{fea06} for solar plages and faculae.
  This result is clearly shown in the right-hand panel of
  Fig.~\ref{fig:eqws_loci}, where we highlight the narrow interval of
  chromospheric pressure within which the majority of the values fall.
  This interval of $P/P_\mathrm{q}$ is mapped in the relatively narrow
  region in the left-hand panel, which empirically corresponds to a
  near-linear joint response of the two \ion{He}{1} lines.

    The linear correlation of the two \ion{He}{1} lines, when computed in
    stars with similar spectral type, is very tight.  For stars with
    $\BV<0.5$, excluding fastest rotators ($\vsini>90$~km~s$^{-1}$: HD~29992
    and HD~189245), the standard deviation around the least-absolute-deviation
    linear fit of the relation \heiv\ vs. \heir\ is of only 3~m\AA.
    The \heiv\ equivalent widths cover a range of 15--40~m\AA\ in this case.
    For stars with $0.5<\BV<0.7$, the standard deviation around the linear
    slope is
    even smaller: about 1~m\AA; the range covered is 5--35~m\AA. 
    The linear fit is relatively less good for
    stars with $\BV>0.7$, with a standard deviation of 4~m\AA\ on smaller
    values of the \dthree\ line than in the other two groups of stars (5--25
    m\AA).  
    There are only four such stars in our sample, however.  

\begin{figure}
\includegraphics[angle=90,width=1.05\linewidth]{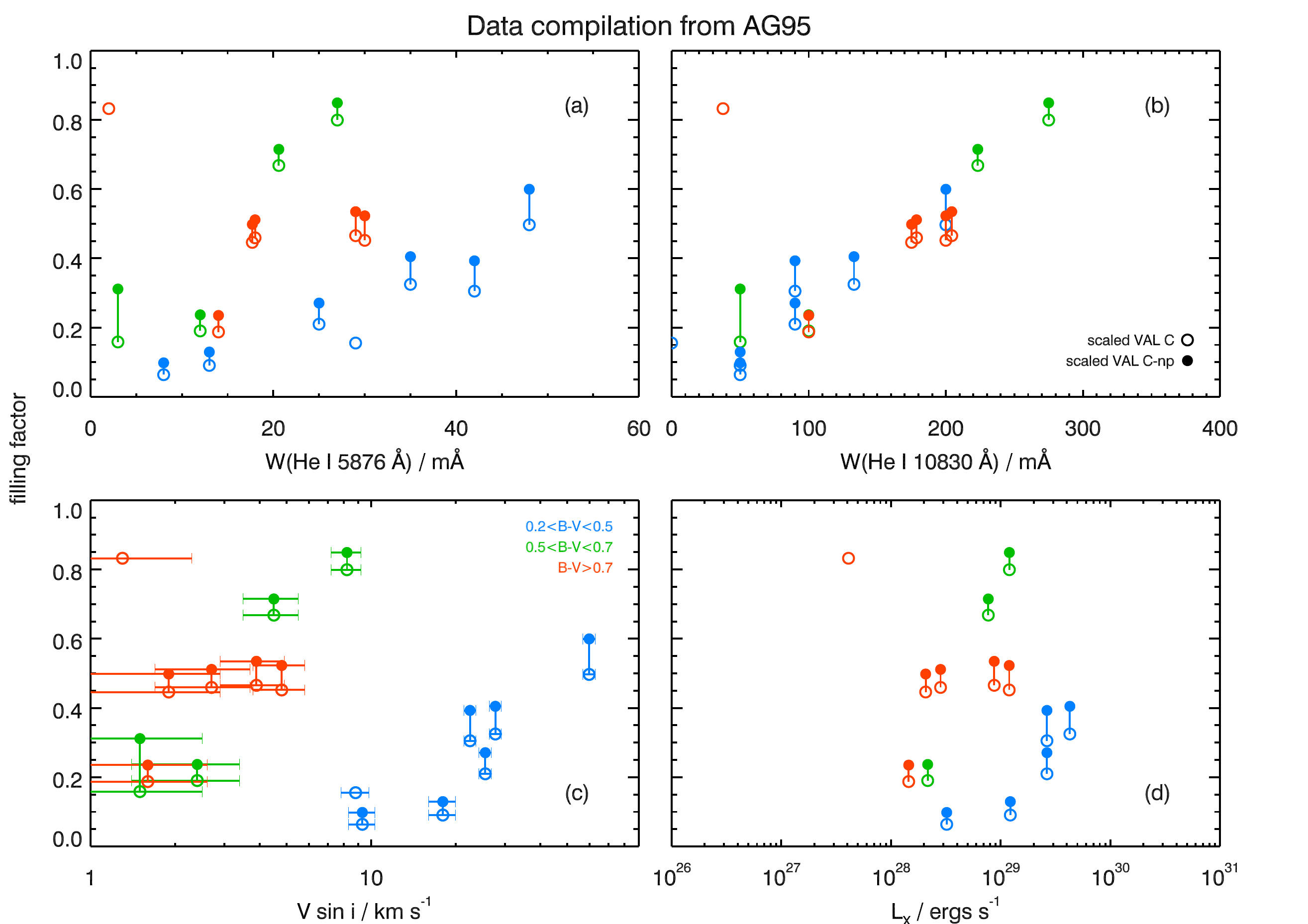}
\caption{%
  Same as Fig.~\ref{fig:eqws_ff}, from a previous
  data compilation of triplet helium lines strengths. 
  Previously acquired measures of
  \heiv\ and \heir\ equivalent widths from AG95 for a different stellar sample
  are added to the data-set discussed herein in order to extend the range of
  the correlation diagrams, in
    particular %
    the correlation with X-ray luminosity.  
    The scatter in the correlation diagrams is larger than in
      Fig.~\ref{fig:eqws_ff} due to the lower quality, non-simultaneity of the
      data. Nevertheless, at least in panel (b) and -- to a lesser extent -- panel
      (a), the linear trends seen in
      that figure appear to be confirmed and extended to lower values of
      \ion{He}{1} equivalent widths.
  The implication is that the intrinsic heating rates in active regions in
  solar-type stars are similar with differences in the observed strength of
  the helium triplet feature due primarily to the area coverage of significant
  concentrations of magnetic flux.  See text for a discussion.
  \label{fig:eqws_extra}
}
\end{figure}

This result would imply that
the principal difference between chromospherically 'active' stars and 'quiet'
solar-type stars is the surface filling factor of active regions.  
This is
only a preliminary conclusion since the range in activity parameters for this
small sample is comparatively narrow as measured by X-ray luminosity.
However, we can add data from AG95 to extend the range, as illustrated in
Fig.~\ref{fig:eqws_extra}.  While these data are non-simultaneous and
generally of lower quality, they nevertheless support the linear trend in
Fig.~\ref{fig:eqws_ff}a and b.            
We also note that that there is evidence
that indeed the variability with the solar cycle of the spectral irradiance
in transition region lines (which are usually correlated with the optical
\ion{He}{1} lines) is mainly driven by the variability of the active region
filling factor, not by changes in the intrinsic contrast with quiescent
regions \citep{adz14}. 

  With regard to the correlation between filling factors and rotational
  velocities, data for stars with $\vsini < 10$~km~s$^{-1}$ shown in
  Fig.~\ref{fig:eqws_extra}c seem to be consistent with the sharp increase
  shown in the corresponding diagram of Fig.~\ref{fig:eqws_ff}c.  In both
  cases, this trend is mostly due to cool stars with $\BV>0.5$.  On the other
  hand, it is less clear whether the decrease of the filling factor with
  \vsini\ above $20$~km~s$^{-1}$ seen Fig.~\ref{fig:eqws_ff}c can be supported
  by these lesser quality, non-simultaneous measures.

  Clearly, an enlarged stellar sample extending over a greater range of
  activity would be desirable before a firm conclusion can be established
  regarding the nature of non-radiative heating in active regions on
  solar-type stars.  In this respect, it is unfortunate we could not complete the
  observations of all our program stars due to adverse weather.  Furthermore,
  extending the solar theoretical models adopted here to a range of effective
  temperatures would help to reduce the uncertainties.

Measurements of these activity
diagnostics at various epochs in the same star could be useful.
But,
if the tentative finding that the morphological differences in activity in
solar-type stars is due primarily to the filling factor of otherwise similar
active regions then this would confirm the basic applicability of the solar
paradigm to such objects.  Moreover, this result is in contrast to what we
find in the case of cooler dwarfs, namely, the M stars that are characterized
by either H$\alpha$ emission (i.e., dMe stars) or {\it chromospheric}
H$\alpha$ absorption.  Since the H$\alpha$ absorption feature in the spectra
of dM (i.e., non-dMe) stars cannot be changed into an emission line simply by
changing the filling factor of active regions alone, the differences between
these stars in the context of magnetic activity must be due to differences in
intrinsic chromospheric heating rates, and, therefore, in the nature of the
magnetic field-related energy dissipation in their atmospheres, at
  least at the level where the line forms.

\section{Summary}
\label{sec:summary}

Our results confirm and extend the previous work of AG95 with a set of
simultaneously acquired spectra of exceptional quality for the principal
objective of estimating the surface coverage of regions of significant
magnetic flux emergence in solar-type stars outside of spots.  We find that
for most of the sample the measured strengths of the \ion{He}{1} triplet
features in Fig.~\ref{fig:eqws_loci} are below the locus of points corresponding to a filling
factor of unity.  In addition, the estimated filling factors given in Fig.~\ref{fig:eqws_ff}
(and the extension in Fig.~\ref{fig:eqws_extra}) suggest that the principal difference between
active and quiet solar-type stars is due to the fractional area coverages
of active regions rather than to a broad range of intrinsic non-radiative
heating rates within active regions. 
  This latter result would very naturally imply a linear relationship between
  the two \ion{He}{1} lines distinct from the optically thin limit and
  parametrized by the filling factor (Eq~\ref{eq:Wobs}).  A linear
  relationship between those two lines whose optical thickness are more than an
  order of magnitude apart would be difficult to explain with radiative
  transfer effects alone, independent of the specific chromospheric models
  adopted.

While the observational results are
broadly consistent with our modeling approach, exceptions are noted.  We
discuss these cases in the context of the need to apply a model atmosphere
that is more appropriate to the stellar type, particularly for the cooler K
dwarfs, and the possibility that non-magnetic contributions along with
observational errors in the warm, very rapidly rotating F dwarfs in our sample
could lead to effects not yet accounted for in the models that we applied.

The {\heir} line seems to be a particularly good indicator of filling
factor. This may be understood from the fact that most of the observations
fall just below the ``flat'' part of the {\heir} vs. {\dthree} curves of joint
responses in all the atmospheric models we investigated.  Hence, the joint
behavior does not discriminate very well between different levels of
``intrinsic'' activity (parameterized by the chromospheric pressure), but the
joint response does discriminate for the filling factor. The various classes
of models only differ in the calibration of the relationship of
$W_{\lambda10830}$ vs. $f$, but all indicate a good linear relationship
between the two quantities.

The main difficulty in measurements of these features arises from the fact
that the \ion{He}{1} triplet lines are relatively weak features in the
spectrum of solar-like stars.  This is exacerbated by the presence of
terrestrial water vapor blends and, in some cases, by stellar blends that require careful corrections, especially
for the {\dthree} line.  Nevertheless, the range of applicability of this
method can be extended to lower, more solar-like levels of activity as well as
enhanced activity levels (typically correlated with rapid rotation) through
high resolution, high signal-to-noise ratio spectra obtained at sites
characterized by very low humidity.

With our approach verified, we intend to obtain spectra for an expanded
stellar sample.  In parallel with the observational effort, we will develop a
grid of chromospheric models as a function of effective temperature that can
be applied to achieve improved discrimination between the possible loci of
filling factor for a range of stellar types.  In principle, the results from this total program could be used to
determine for the very first time the empirical relationship between rotation
and active region area coverage. Indeed, while correlations between rotation
and radiative proxies of magnetic flux, such as \ion{Ca}{2} or X-rays, have
been determined for various stellar samples, the empirical relationship
between actual magnetic active region area coverage and rotation has not been
established.

Methods for independent verification of inferred active region filling factors
may be pursued through other techniques such as Doppler imaging or by
observations of eclipsing binaries.  One novel approach is to obtain spectra
of the helium triplet features during occultations of surface active regions
by transiting exoplanets, specifically for systems with super-Jupiters in
proximity to an active host star.  In particular, signatures of exoplanet
occultations of starspots have been observed with the high-precision
photometry from the CoRot and \textit{Kepler} missions \citep{wol09,fra14}.
Analogously, distortions of the line profile in the form of relative increases
or ``bumps'' should be seen in the triplet features in high-quality,
time-resolved spectra, in analogy with spectroscopic signatures of
  transiting exoplanets described in, e.~g. \cite{mancini-etal:15}.  

\acknowledgments

This work is based on observations collected at the European Organisation for
Astronomical Research in the Southern Hemisphere  %
with ESO Telescopes at the La Silla Paranal Observatory  %
under program ID 088.D-0028(A) and MPG Utility Run for FEROS 088.A-9029(A).
We are grateful to the ESO TAC and the Observatory technical support
  staff for supporting this program.
We also thank prof.\ Manfred Sch\"{u}ssler for valuable discussions in the
definition phase of this project. 
Finally, we thank an anonymous referee for a
thorough reading of the manuscript and for 
comments that served to enhance the quality of presentation.

\vspace{5mm}
\facilities{ESO VLT (CRIRES), %
  ESO MPG (FEROS), %
  NSF McMath-Pierce (FTS), %
  NSF SOLIS (VSM), %
  NASA SDO (AIA)}

\software{MIDAS, %
  ESOREX, %
  IDL}


\begin{thebibliography}{}
\expandafter\ifx\csname natexlab\endcsname\relax\def\natexlab#1{#1}\fi

\bibitem[{{Ammler-von Eiff} \& {Reiners}(2012)}]{vsini12}
{Ammler-von Eiff}, M., \& {Reiners}, A. 2012, \aap, 542, A116

\bibitem[{{Andretta}(1994 (A94))}]{a94}
{Andretta}, V. 1994 (A94), PhD thesis, Univ.~Naples ``Federico II''

\bibitem[{{Andretta} \& {Del Zanna}(2014)}]{adz14}
{Andretta}, V., \& {Del Zanna}, G. 2014, \aap, 563, A26

\bibitem[{{Andretta} {et~al.}(2003){Andretta}, {Del Zanna}, \&
  {Jordan}}]{andretta-etal:03}
{Andretta}, V., {Del Zanna}, G., \& {Jordan}, S.~D. 2003, \aap, 400, 737

\bibitem[{{Andretta} \& {Giampapa}(1995 (AG95))}]{ag95}
{Andretta}, V., \& {Giampapa}, M.~S. 1995 (AG95), \apj, 439, 405

\bibitem[{{Andretta} {et~al.}(1995){Andretta}, {Giampapa}, \& {Jones}}]{agj95}
{Andretta}, V., {Giampapa}, M.~S., \& {Jones}, H.~P. 1995, Irish Astronomical
  Journal, 22, 177

\bibitem[{{Andretta} \& {Jones}(1997 (AJ97))}]{aj97}
{Andretta}, V., \& {Jones}, H.~P. 1997 (AJ97), \apj, 489, 375

\bibitem[{{Andretta} {et~al.}(2000){Andretta}, {Jordan}, {Brosius}, {Davila},
  {Thomas}, {Behring}, {Thompson}, \& {Garcia}}]{andretta-etal:00}
{Andretta}, V., {Jordan}, S.~D., {Brosius}, J.~W., {et~al.} 2000, \apj, 535,
  438

\bibitem[{{Andretta} {et~al.}(2008){Andretta}, {Mauas}, {Falchi}, \&
  {Teriaca}}]{andretta-etal:08}
{Andretta}, V., {Mauas}, P.~J.~D., {Falchi}, A., \& {Teriaca}, L. 2008, \apj,
  681, 650

\bibitem[{{Baglin} {et~al.}(2009){Baglin}, {Auvergne}, {Barge}, {Deleuil},
  {Michel}, \& {CoRoT Exoplanet Science Team}}]{Baglin-etal:09}
{Baglin}, A., {Auvergne}, M., {Barge}, P., {et~al.} 2009, in IAU Symposium,
  Vol. 253, Transiting Planets, ed. F.~{Pont}, D.~{Sasselov}, \& M.~J.
  {Holman}, 71--81

\bibitem[{{Baliunas} {et~al.}(1995){Baliunas}, {Donahue}, {Soon}, {Horne},
  {Frazer}, {Woodard-Eklund}, {Bradford}, {Rao}, {Wilson}, {Zhang}, {Bennett},
  {Briggs}, {Carroll}, {Duncan}, {Figueroa}, {Lanning}, {Misch}, {Mueller},
  {Noyes}, {Poppe}, {Porter}, {Robinson}, {Russell}, {Shelton}, {Soyumer},
  {Vaughan}, \& {Whitney}}]{bali95}
{Baliunas}, S.~L., {Donahue}, R.~A., {Soon}, W.~H., {et~al.} 1995, \apj, 438,
  269

\bibitem[{{Brault}(1979)}]{Brault79}
{Brault}, J.~W. 1979, Osservazioni e Memorie dell'Osservatorio Astrofisico di
  Arcetri, 106, 33

\bibitem[{{Breckinridge} \& {Hall}(1973)}]{breckinridge-hall:73}
{Breckinridge}, J.~B., \& {Hall}, D.~N.~B. 1973, \solphys, 28, 15

\bibitem[{{Buzasi} {et~al.}(2016){Buzasi}, {Lezcano}, \&
  {Preston}}]{Buzasi-etal:16}
{Buzasi}, D., {Lezcano}, A., \& {Preston}, H.~L. 2016, Journal of Space Weather
  and Space Climate, 6, A38

\bibitem[{{Cram} \& {Mullan}(1979)}]{cm79}
{Cram}, L.~E., \& {Mullan}, D.~J. 1979, \apj, 234, 579

\bibitem[{{Dupree} {et~al.}(2012){Dupree}, {Brickhouse}, {Cranmer}, {Luna},
  {Schneider}, {Bessell}, {Bonanos}, {Crause}, {Lawson}, {Mallik}, \&
  {Schuler}}]{Dupree-etal:12}
{Dupree}, A.~K., {Brickhouse}, N.~S., {Cranmer}, S.~R., {et~al.} 2012, \apj,
  750, 73

\bibitem[{{Fontenla} {et~al.}(2006){Fontenla}, {Avrett}, {Thuillier}, \&
  {Harder}}]{fea06}
{Fontenla}, J.~M., {Avrett}, E., {Thuillier}, G., \& {Harder}, J. 2006, \apj,
  639, 441

\bibitem[{{Fontenla} {et~al.}(1993){Fontenla}, {Avrett}, \& {Loeser}}]{fal93}
{Fontenla}, J.~M., {Avrett}, E.~H., \& {Loeser}, R. 1993, \apj, 406, 319

\bibitem[{{Fraine} {et~al.}(2014){Fraine}, {Deming}, {Benneke}, {Knutson},
  {Jord{\'a}n}, {Espinoza}, {Madhusudhan}, {Wilkins}, \& {Todorov}}]{fra14}
{Fraine}, J., {Deming}, D., {Benneke}, B., {et~al.} 2014, \nat, 513, 526

\bibitem[{{Garc\'\i a-L\'opez} {et~al.}(1993){Garc\'\i a-L\'opez}, {Rebolo},
  {Beckman}, \& {McKeith}}]{garcialopez-etal:93}
{Garc\'\i a-L\'opez}, R.~J., {Rebolo}, R., {Beckman}, J.~E., \& {McKeith},
  C.~D. 1993, \aap, 273, 482

\bibitem[{{Giampapa}(1985)}]{giam85}
{Giampapa}, M.~S. 1985, \apj, 299, 781

\bibitem[{{Giampapa} {et~al.}(1982){Giampapa}, {Worden}, \& {Linsky}}]{giam82}
{Giampapa}, M.~S., {Worden}, S.~P., \& {Linsky}, J.~L. 1982, \apj, 258, 740

\bibitem[{{Giovanelli} \& {Hall}(1977)}]{giovanelli-hall:77}
{Giovanelli}, R.~G., \& {Hall}, D. 1977, \solphys, 52, 211

\bibitem[{{G{\l}{\c e}bocki} \& {Gnaci{\'n}ski}(2005)}]{vsini05}
{G{\l}{\c e}bocki}, R., \& {Gnaci{\'n}ski}, P. 2005, {Catalog of Stellar
  Rotational Velocities},  {VizieR Online Data Catalog III/244}

\bibitem[{{Gliese} \& {Jahreiss}(1991)}]{gj95}
{Gliese}, W., \& {Jahreiss}, H. 1991, {Preliminary Version of the Third
  Catalogue of Nearby Stars (CNS3)}, ed. L.~Brotzmann \& S.~Gesser, Vol.~I
  (Greenbelt MD: NASA/Astronomical Data Center, GSFC)

\bibitem[{{Harvey} {et~al.}(1975){Harvey}, {Krieger}, {Timothy}, \&
  {Vaiana}}]{jharvey75}
{Harvey}, J., {Krieger}, A.~S., {Timothy}, A.~F., \& {Vaiana}, G.~S. 1975,
  Osservazioni e Memorie dell'Osservatorio Astrofisico di Arcetri, 104, 50

\bibitem[{{Harvey} \& {Livingston}(1994)}]{harvey-livingston:94}
{Harvey}, J.~W., \& {Livingston}, W.~C. 1994, in IAU Symposium, Vol. 154,
  Infrared Solar Physics, ed. D.~M. {Rabin}, J.~T. {Jefferies}, \&
  C.~{Lindsey}, 59

\bibitem[{{He} {et~al.}(2015){He}, {Wang}, \& {Yun}}]{He-etal:15}
{He}, H., {Wang}, H., \& {Yun}, D. 2015, \apjs, 221, 18

\bibitem[{{Hoffleit} \& {Jaschek}(1991)}]{bsc91}
{Hoffleit}, D., \& {Jaschek}, C. 1991, {The Bright star catalogue}, {5th} edn.,
  ed. D.~Hoffleit \& C.~Jaschek ({New Haven, CT}: {Yale Univ. Obs.})

\bibitem[{{H{\"u}nsch} {et~al.}(1999){H{\"u}nsch}, {Schmitt}, {Sterzik}, \&
  {Voges}}]{ROSATnearby99}
{H{\"u}nsch}, M., {Schmitt}, J.~H.~M.~M., {Sterzik}, M.~F., \& {Voges}, W.
  1999, \aaps, 135, 319

\bibitem[{{H{\"u}nsch} {et~al.}(1998){H{\"u}nsch}, {Schmitt}, \&
  {Voges}}]{ROSATbright98}
{H{\"u}nsch}, M., {Schmitt}, J.~H.~M.~M., \& {Voges}, W. 1998, \aaps, 132, 155

\bibitem[{{Jones}(2003)}]{jones:03}
{Jones}, H.~P. 2003, \solphys, 218, 1

\bibitem[{{Judge} \& {Pietarila}(2004)}]{judge-pietarila:04}
{Judge}, P.~G., \& {Pietarila}, A. 2004, \apj, 606, 1258

\bibitem[{{Kaufer} {et~al.}(2000){Kaufer}, {Stahl}, {Tubbesing}, {Norregaard},
  {Avila}, {Francois}, {Pasquini}, \& {Pizzella}}]{kaufer-etal:00}
{Kaufer}, A., {Stahl}, O., {Tubbesing}, S., {et~al.} 2000, in SPIE Conference
  Series, Vol. 4008, Optical and IR Telescope Instrumentation and Detectors,
  ed. M.~{Iye} \& A.~F. {Moorwood}, 459--466

\bibitem[{{K{\"a}ufl} {et~al.}(2004){K{\"a}ufl}, {Ballester}, {Biereichel},
  {Delabre}, {Donaldson}, {Dorn}, {Fedrigo}, {Finger}, {Fischer}, {Franza},
  {Gojak}, {Huster}, {Jung}, {Lizon}, {Mehrgan}, {Meyer}, {Moorwood}, {Pirard},
  {Paufique}, {Pozna}, {Siebenmorgen}, {Silber}, {Stegmeier}, \&
  {Wegerer}}]{crires04}
{K{\"a}ufl}, H.-U., {Ballester}, P., {Biereichel}, P., {et~al.} 2004, in SPIE
  Conference Series, Vol. 5492, Ground-based Instrumentation for Astronomy, ed.
  A.~F.~M. {Moorwood} \& M.~{Iye}, 1218--1227

\bibitem[{{K{\"a}ufl} {et~al.}(2006){K{\"a}ufl}, {Amico}, {Ballester},
  {Bendek}, {Biereichel}, {Bristow}, {Casali}, {Delabre}, {Dorn}, {Eschbaumer},
  {Esteves}, {Fedrigo}, {Finger}, {Fischer}, {Gillet}, {Gojak}, {Huster},
  {Jung}, {Kerber}, {Kirchbauer}, {Lizon}, {Marchetti}, {Mehrgan}, {Meyer},
  {Moorwood}, {Oberti}, {Pirard}, {Paufique}, {Pozna}, {Primas}, {Schmutzer},
  {Seifahrt}, {Siebenmorgen}, {Silber}, {Smette}, {Sokar}, {Stegmeier},
  {Tacconi-Garman}, {Tordo}, {Uttenthaler}, \& {Weilenmann}}]{crires06}
{K{\"a}ufl}, H.~U., {Amico}, P., {Ballester}, P., {et~al.} 2006, The Messenger,
  126, 32

\bibitem[{{Koch} {et~al.}(2010){Koch}, {Borucki}, {Basri}, {Batalha}, {Brown},
  {Caldwell}, {Christensen-Dalsgaard}, {Cochran}, {DeVore}, {Dunham},
  {Gautier}, {Geary}, {Gilliland}, {Gould}, {Jenkins}, {Kondo}, {Latham},
  {Lissauer}, {Marcy}, {Monet}, {Sasselov}, {Boss}, {Brownlee}, {Caldwell},
  {Dupree}, {Howell}, {Kjeldsen}, {Meibom}, {Morrison}, {Owen}, {Reitsema},
  {Tarter}, {Bryson}, {Dotson}, {Gazis}, {Haas}, {Kolodziejczak}, {Rowe}, {Van
  Cleve}, {Allen}, {Chandrasekaran}, {Clarke}, {Li}, {Quintana}, {Tenenbaum},
  {Twicken}, \& {Wu}}]{Koch-etal:10}
{Koch}, D.~G., {Borucki}, W.~J., {Basri}, G., {et~al.} 2010, \apjl, 713, L79

\bibitem[{{Kurucz} {et~al.}(1984){Kurucz}, {Furenlid}, {Brault}, \&
  {Testerman}}]{kurucz84}
{Kurucz}, R.~L., {Furenlid}, I., {Brault}, J., \& {Testerman}, L. 1984, {Solar
  flux atlas from 296 to 1300 nm} (Sunspot, NM: National Solar Observatory)

\bibitem[{{Lambert} \& {O'Brien}(1983)}]{lambert-obrien:83}
{Lambert}, D.~L., \& {O'Brien}, G.~T. 1983, \aap, 128, 110

\bibitem[{{Landman}(1981)}]{Landman81}
{Landman}, D.~A. 1981, \apj, 244, 345

\bibitem[{{Leenaarts} {et~al.}(2016){Leenaarts}, {Golding}, {Carlsson},
  {Libbrecht}, \& {Joshi}}]{Leenaarts-etal:16}
{Leenaarts}, J., {Golding}, T., {Carlsson}, M., {Libbrecht}, T., \& {Joshi}, J.
  2016, \aap, 594, A104

\bibitem[{{Livingston} {et~al.}(2010){Livingston}, {White}, {Wallace}, \&
  {Harvey}}]{livingston-etal:10}
{Livingston}, W., {White}, O.~R., {Wallace}, L., \& {Harvey}, J. 2010, \memsai,
  81, 643

\bibitem[{{MacPherson} \& {Jordan}(1999)}]{macpherson-jordan:99}
{MacPherson}, K.~P., \& {Jordan}, C. 1999, \mnras, 308, 510

\bibitem[{{Malanushenko} \& {Jones}(2004)}]{malanushenko-jones:04}
{Malanushenko}, O.~V., \& {Jones}, H.~P. 2004, \solphys, 222, 43

\bibitem[{{Mancini} {et~al.}(2015){Mancini}, {Esposito}, {Covino}, {Raia},
  {Southworth}, {Tregloan-Reed}, {Biazzo}, {Bonomo}, {Desidera}, {Lanza},
  {Maciejewski}, {Poretti}, {Sozzetti}, {Borsa}, {Bruni}, {Ciceri}, {Claudi},
  {Cosentino}, {Gratton}, {Martinez Fiorenzano}, {Lodato}, {Lorenzi},
  {Marzari}, {Murabito}, {Affer}, {Bignamini}, {Bedin}, {Boccato}, {Damasso},
  {Henning}, {Maggio}, {Micela}, {Molinari}, {Pagano}, {Piotto}, {Rainer},
  {Scandariato}, {Smareglia}, \& {Zanmar Sanchez}}]{mancini-etal:15}
{Mancini}, L., {Esposito}, M., {Covino}, E., {et~al.} 2015, \aap, 579, A136

\bibitem[{{Mauas} {et~al.}(2005){Mauas}, {Andretta}, {Falchi}, {Falciani},
  {Teriaca}, \& {Cauzzi}}]{mauas-etal:05}
{Mauas}, P.~J.~D., {Andretta}, V., {Falchi}, A., {et~al.} 2005, \apj, 619, 604

\bibitem[{{McQuillan} {et~al.}(2014){McQuillan}, {Mazeh}, \&
  {Aigrain}}]{McQuillan-etal:14}
{McQuillan}, A., {Mazeh}, T., \& {Aigrain}, S. 2014, \apjs, 211, 24

\bibitem[{{Moore} {et~al.}(1966){Moore}, {Minnaert}, \&
  {Houtgast}}]{moore-etal:66}
{Moore}, C.~E., {Minnaert}, M.~G.~J., \& {Houtgast}, J. 1966, {The solar
  spectrum 2935 A to 8770 A, National Bureau of Standards Monograph 61}
  (Washington, DC: US Government Printing Office)

\bibitem[{{Muglach} \& {Schmidt}(2001)}]{muglach-schmidt:01}
{Muglach}, K., \& {Schmidt}, W. 2001, \aap, 379, 592

\bibitem[{{Nielsen} {et~al.}(2013){Nielsen}, {Gizon}, {Schunker}, \&
  {Karoff}}]{Nielsen-etal:13}
{Nielsen}, M.~B., {Gizon}, L., {Schunker}, H., \& {Karoff}, C. 2013, \aap, 557,
  L10

\bibitem[{{Peres} {et~al.}(2000){Peres}, {Orlando}, {Reale}, {Rosner}, \&
  {Hudson}}]{peres-etal:00}
{Peres}, G., {Orlando}, S., {Reale}, F., {Rosner}, R., \& {Hudson}, H. 2000,
  \apj, 528, 537

\bibitem[{{Pietarila} \& {Judge}(2004)}]{pietarila-judge:04}
{Pietarila}, A., \& {Judge}, P.~G. 2004, \apj, 606, 1239

\bibitem[{{Pizzolato} {et~al.}(2003){Pizzolato}, {Maggio}, {Micela},
  {Sciortino}, \& {Ventura}}]{pizzolato-etal:03}
{Pizzolato}, N., {Maggio}, A., {Micela}, G., {Sciortino}, S., \& {Ventura}, P.
  2003, \aap, 397, 147

\bibitem[{{Reinhold} {et~al.}(2013){Reinhold}, {Reiners}, \&
  {Basri}}]{Reinhold-etal:13}
{Reinhold}, T., {Reiners}, A., \& {Basri}, G. 2013, \aap, 560, A4

\bibitem[{{Saar} {et~al.}(1997){Saar}, {Huovelin}, {Osten}, \&
  {Shcherbakov}}]{saar-etal:97}
{Saar}, S.~H., {Huovelin}, J., {Osten}, R.~A., \& {Shcherbakov}, A.~G. 1997,
  \aap, 326, 741

\bibitem[{{Santos} {et~al.}(2010){Santos}, {Gomes da Silva}, {Lovis}, \&
  {Melo}}]{santos-etal:10}
{Santos}, N.~C., {Gomes da Silva}, J., {Lovis}, C., \& {Melo}, C. 2010, \aap,
  511, A54

\bibitem[{{Sanz-Forcada} \& {Dupree}(2008)}]{SanzForcada-Dupree:08}
{Sanz-Forcada}, J., \& {Dupree}, A.~K. 2008, \aap, 488, 715

\bibitem[{{Schisano} {et~al.}(2009){Schisano}, {Covino}, {Alcal{\'a}},
  {Esposito}, {Gandolfi}, \& {Guenther}}]{schisano-etal:09}
{Schisano}, E., {Covino}, E., {Alcal{\'a}}, J.~M., {et~al.} 2009, \aap, 501,
  1013

\bibitem[{{Schr{\"o}der} {et~al.}(2009){Schr{\"o}der}, {Reiners}, \&
  {Schmitt}}]{vsini09}
{Schr{\"o}der}, C., {Reiners}, A., \& {Schmitt}, J.~H.~M.~M. 2009, \aap, 493,
  1099

\bibitem[{{Skumanich} {et~al.}(1975){Skumanich}, {Smythe}, \&
  {Frazier}}]{sku75}
{Skumanich}, A., {Smythe}, C., \& {Frazier}, E.~N. 1975, \apj, 200, 747

\bibitem[{{Smith} {et~al.}(2012{\natexlab{a}}){Smith}, {Dupree}, \&
  {Strader}}]{GSmith-etal:12}
{Smith}, G.~H., {Dupree}, A.~K., \& {Strader}, J. 2012{\natexlab{a}}, \pasp,
  124, 1252

\bibitem[{{Smith} {et~al.}(2012{\natexlab{b}}){Smith}, {Dupree}, \&
  {Strader}}]{smith-etal:12}
---. 2012{\natexlab{b}}, \pasp, 124, 1252

\bibitem[{{Smith}(2003)}]{smith:03}
{Smith}, G.~R. 2003, \mnras, 341, 143

\bibitem[{{Smith} \& {Jordan}(2002)}]{smith-jordan:02}
{Smith}, G.~R., \& {Jordan}, C. 2002, \mnras, 337, 666

\bibitem[{{Swensson} {et~al.}(1970){Swensson}, {Benedict}, {Delbouille}, \&
  {Roland}}]{swensson-etal:70}
{Swensson}, J.~W., {Benedict}, W.~S., {Delbouille}, L., \& {Roland}, G. 1970,
  Memoires of the Societe Royale des Sciences de Liege, 5

\bibitem[{{Takeda} \& {Takada-Hidai}(2011)}]{takeda-takadahidai:11}
{Takeda}, Y., \& {Takada-Hidai}, M. 2011, \pasj, 63, 547

\bibitem[{{Tobias}(1997)}]{tobias97}
{Tobias}, S.~M. 1997, \aap, 322, 1007

\bibitem[{{Tu} \& {Song}(2013)}]{tu13}
{Tu}, J., \& {Song}, P. 2013, \apj, 777, 53

\bibitem[{{Vernazza} {et~al.}(1981){Vernazza}, {Avrett}, \& {Loeser}}]{val81}
{Vernazza}, J.~E., {Avrett}, E.~H., \& {Loeser}, R. 1981, \apjs, 45, 635

\bibitem[{{Voges} {et~al.}(1999){Voges}, {Aschenbach}, {Boller},
  {Br{\"a}uninger}, {Briel}, {Burkert}, {Dennerl}, {Englhauser}, {Gruber},
  {Haberl}, {Hartner}, {Hasinger}, {K{\"u}rster}, {Pfeffermann}, {Pietsch},
  {Predehl}, {Rosso}, {Schmitt}, {Tr{\"u}mper}, \& {Zimmermann}}]{voges99}
{Voges}, W., {Aschenbach}, B., {Boller}, T., {et~al.} 1999, \aap, 349, 389

\bibitem[{{Wilson}(1978)}]{wilson78}
{Wilson}, O.~C. 1978, \apj, 226, 379

\bibitem[{{Wolff} {et~al.}(1986){Wolff}, {Boesgaard}, \&
  {Simon}}]{wolff-etal:86}
{Wolff}, S.~C., {Boesgaard}, A.~M., \& {Simon}, T. 1986, \apj, 310, 360

\bibitem[{{Wolff} \& {Heasley}(1984)}]{wolff-heasley:84}
{Wolff}, S.~C., \& {Heasley}, J.~N. 1984, \pasp, 96, 231

\bibitem[{{Wolff} \& {Heasley}(1987)}]{wolff-heasley:87}
---. 1987, \pasp, 99, 957

\bibitem[{{Wolff} {et~al.}(1985){Wolff}, {Heasley}, \&
  {Varsik}}]{wolff-etal:85}
{Wolff}, S.~C., {Heasley}, J.~N., \& {Varsik}, J. 1985, \pasp, 97, 707

\bibitem[{{Wolter} {et~al.}(2009){Wolter}, {Schmitt}, {Huber}, {Czesla},
  {M{\"u}ller}, {Guenther}, \& {Hatzes}}]{wol09}
{Wolter}, U., {Schmitt}, J.~H.~M.~M., {Huber}, K.~F., {et~al.} 2009, \aap, 504,
  561

\bibitem[{{Yeo} {et~al.}(2014){Yeo}, {Krivova}, \& {Solanki}}]{yeo-etal:14}
{Yeo}, K.~L., {Krivova}, N.~A., \& {Solanki}, S.~K. 2014, \ssr, 186, 137

\bibitem[{{Zarro} \& {Zirin}(1986)}]{Zarro-Zirin:86}
{Zarro}, D.~M., \& {Zirin}, H. 1986, \apj, 304, 365

\end{thebibliography}
\end{document}